\documentclass[11pt]{article}
\usepackage{geometry,float}
\usepackage[parfill]{parskip}
\usepackage{amsmath,amssymb,bbm}
\usepackage{cases}
\usepackage{graphicx,psfrag,epsf}
\usepackage{caption}
\usepackage{subcaption}
\usepackage{tabularx}
\usepackage{microtype}
\usepackage{enumitem}
\usepackage{authblk}
\usepackage{url}
\usepackage{lineno}
\usepackage{tikz}
\usepackage[colorlinks,citecolor=blue,urlcolor=blue,linkcolor=blue,linktocpage=true]{hyperref}
\usepackage[amsmath,amsthm,thmmarks]{ntheorem}
\usepackage{cleveref}
\crefformat{equation}{(#2#1#3)}
\crefrangeformat{equation}{(#3#1#4) to~(#5#2#6)}
\crefname{equation}{}{}
\Crefname{equation}{}{}
\usepackage[authoryear]{natbib}

\newtheorem{theorem}{Theorem}

\newtheorem{assumption}{Assumption}
\newtheorem{algorithm}{Algorithm}
\newtheorem{definition}{Definition}

\let\hat\widehat

\newcommand{\fdiv}{\mathbbmss f}

\newcommand{\overbar}[1]{\mkern 1.5mu\overline{\mkern-1.5mu#1\mkern-1.5mu}\mkern 1.5mu}

\renewenvironment{proof}{{\bf Proof }}{$\Box$}

\usepackage{color}

\begin{document}

\def\spacingset#1{\renewcommand{\baselinestretch}%
{#1}\small\normalsize} \spacingset{1}

\title{\bf \Large{Sequentially additive nonignorable missing data modeling using auxiliary marginal information}} 
\author{Mauricio Sadinle and Jerome P. Reiter}
		\date{University of Washington and Duke University}
  \maketitle

\begin{abstract}
We study a class of missingness mechanisms, called sequentially additive nonignorable, for modeling multivariate data with item nonresponse. These mechanisms explicitly allow the probability of nonresponse for each variable to depend on the value of that variable, thereby representing  nonignorable missingness mechanisms.  These missing data models are identified by making use of auxiliary information on marginal distributions, such as marginal probabilities for multivariate categorical variables or moments for numeric variables.  We present theory proving identification results, and illustrate the use of these mechanisms in an application.
\end{abstract}

\noindent%

\spacingset{1.4} 
\section{Introduction}

When data values are unintentionally missing, analysts generally cannot estimate the true probabilistic mechanism generating the missingness from the observed data alone. To proceed with statistical inference, they have to make
unverifiable assumptions on how the missingness arises. These identifying assumptions correspond to restrictions on the joint distribution of the study variables and their missingness indicators.  
Often in practice, analysts specify such restrictions on the conditional distribution of the missingness indicators given the study variables; for example, they assume the data are
missing at random \citep{Rubin76}. 
In many contexts, it is desirable to use restrictions that encode nonignorable missingness mechanisms 
\citep{Greenleesetal82,Robins97,SadinleReiter17}. 
However, for multivariate data with arbitrary patterns of nonresponse, it can be difficult to specify such nonignorable mechanisms in ways that lead to provably identifiable models \citep[][Section 5.9.1]{ibrahim, DanielsHogan08}.

In this article, we study a class of such missingness mechanisms, called sequentially additive nonignorable, for handling item nonresponse in multiple variables.  We adopt this terminology from \cite{HoonhoutRidder18}, who introduced the sequentially additive nonignorable attrition mechanism for longitudinal studies with monotone nonresponse, that is, when participants who drop out no longer return to the study.  
Similarly as with the attrition mechanism, the missingness mechanism explicitly allows item nonresponse in multiple variables to depend on the values of those variables.  We prove that identifiability is attainable using different types of auxiliary marginal information, that is, external information on features of the distribution of the study variables.  Examples of external data sources include censuses and administrative databases, which readily provide the population distribution of some variables; large surveys, which provide high-quality estimates of some population characteristics; and refreshment samples in longitudinal studies, where at later waves additional observations are drawn from the population and whose responses are fully recorded \citep{Hiranoetal98,Hiranoetal01,Dengetal13,HoonhoutRidder18}. 

The sequentially additive nonignorable missingness mechanism builds on the additive nonignorable mechanism of \cite{Hiranoetal98,Hiranoetal01} and on the attrition mechanism of \cite{HoonhoutRidder18}, with three key differences.  First and most critically, \cite{Hiranoetal98,Hiranoetal01} focused on the case of one variable subject to missingness, and \cite{HoonhoutRidder18} on multiple variables subject to a monotone missingness pattern, whereas we consider the more general case of multiple variables under nonmonotone nonresponse.  Second, their identification results apply to either exclusively categorical or exclusively continuous variables, whereas our results apply to general variable types in arbitrary probability spaces.   Third, our identification results can be used with different types of auxiliary marginal information, such as moments of random variables, or univariate and multivariate marginal distributions, unlike \cite{Hiranoetal98,Hiranoetal01} and \cite{HoonhoutRidder18} which require univariate marginal distributions.

\section{Set-up and preliminaries}\label{s:setup}

\subsection{Notation}\label{notation}

Let the random vectors $Y=(Y_1,\ldots,Y_p)$ and $X=(X_1,\ldots,X_q)$ represent the study variables, where auxiliary marginal information on the distribution of $Y$ is available.  We formally define auxiliary marginal information in Section \ref{ss:ami}.  In general, the variables in both $X$ and $Y$ might be subject to missingness.  Let $M=(M_1,\ldots,M_p)$ be the vector of missingness indicators for $Y$, where $M_j=1$ when $Y_j$ is missing and $M_j=0$ when $Y_j$ is observed.  The vector $W=(W_1,\ldots,W_q)$ of missingness indicators for $X$ is defined analogously.  We  denote $Y_{<j}=(Y_1,\ldots,Y_{j-1})$, $Y_{\leq j}=(Y_1,\ldots,Y_{j})$, $Y_{>j}=(Y_{j+1},\ldots,Y_{p})$, $Y_{\geq j}=(Y_{j},\ldots,Y_{p})$, and similarly for subvectors of $X$, $W$ and $M$.  We denote generic possible values of $Y$ and $Y_{< j}$ by $y$ and $y_{< j}$, respectively, and similarly for other random vectors.  

Let $\mu_j$ and $\nu_l$ be dominating measures for the marginal distributions of $Y_j$ and $X_l$, respectively, and let $\mu=\otimes_{j=1}^p\mu_j$ and $\nu=\otimes_{l=1}^q\nu_l$.  The full-data distribution is the joint distribution of $(X,Y,W,M)$, and we refer to its density $f(x,y,w,m)$ with respect to the product of $\nu$, $\mu$, and the counting measure on $\{0,1\}^{p+q}$ as the full-data density.  The conditional distribution with density $f(w,m \mid x,y)$ is referred to as the missingness mechanism \citep[][p. 90]{DanielsHogan08}.  For simplicity we use $f$ for technically different functions, but their actual interpretations should be clear from the arguments passed to them.  For example, we denote the density of the marginal distribution of $Y$ by $f(y)$.  
The set of functions $u$ such that $\int |u(y)| f(y)\mu(dy)<\infty$ is denoted $L_1\{f(y)\}$, and similarly for other densities.  The linear span of a set of functions $\mathcal{U}$ is denoted $\langle\mathcal{U}\rangle$, and its closure $\overbar{\langle\mathcal{U}\rangle}$.

A missingness pattern for the $Y$ variables is represented by $m=(m_1,\ldots,m_p)\in \{0,1\}^p$.  Given $m$ we 
define $\bar{m}=1_p-m$ to be the indicator vector of observed $Y$ variables, where $1_p$ 
is a vector of ones of length $p$.  We define $Y_{m}=(Y_j: m_j=1)$ to be the missing $Y$ variables and $Y_{\bar{m}}=(Y_j: \bar m_j=1)$ to be the observed $Y$ variables according to $m$. 
We define $w=(w_1,\ldots,w_q)\in \{0,1\}^q$, $\bar{w}$, $X_{w}$ and $X_{\bar w}$ in an analogous fashion.  
The observed-data distribution is the distribution involving the missingness indicators and the corresponding observed variables, with density  
\begin{equation}\label{eq:comp_obsdata}
f(x_{\bar{w}},y_{\bar{m}},w,m)=\iint f(x,y,w,m)\nu(dx_{w})\mu(dy_{m}).
\end{equation}

A more compact way of representing the observed-data distribution is obtained by introducing what \cite{SadinleReiter18} call materialized variables, defined as 
\begin{equation*}
Y_j^* \equiv \left\{\begin{array}{cc}
        Y_j, & \text{if } M_j=0;\\
        *, & \text{if } M_j=1;
        \end{array}\right.
\end{equation*}
where $*$ is simply a placeholder for missingness \citep{Rubin76}.  We define analogously $X^*_l$, and denote $Y^*=(Y_1^*,\ldots,Y_p^*)$ and $X^*=(X_1^*,\ldots,X_q^*)$.  All the observable information from the sampling process can be obtained from the materialized variables:  if $Y_j^*=*$ then $Y_j$ is not observed, and if $Y_j^*=y_j\neq *$ then $Y_j$ is observed and $Y_j=y_j$.  This means that the distribution of $(X^*,Y^*)$ is nothing but a different way of representing the observed-data distribution.  
Therefore, with some abuse of notation,  
the observed-data density can be written in terms of  $X^*$ and $Y^*$, that is, $f(x^*,y^*)\equiv f(x_{\bar{w}},y_{\bar{m}},w,m),$ where $y^*=(y^*_1,\ldots,y^*_p)$ with $y^*_j=*$ if $m_j=1$ and $y^*_j=y_j$ if  $m_j=0$, likewise for $x^*$.  We also often use a hybrid notation as in $f(x,y^*)\equiv f(x,y_{\bar{m}},m)$.

 For our identification results we will work with generic random vectors $(X^*,Y^*)$, but for estimation we will assume to have observations that are drawn as independent and identically distributed copies of $(X^*,Y^*)$, as explained in Section \ref{ss:likelihood}.

\subsection{Identifiability}\label{identifiable}

The full-data distribution cannot be identified from observed data alone in a nonparametric manner; even if we had the ability to sample indefinitely we would only be able to recover the observed-data distribution.  
This fact leads us to introduce some necessary definitions.   We first present the concept of observational equivalence, which we borrow from \cite{Koopmans49}.

\begin{definition}[Observational equivalence] Two full-data distributions are said to be observationally equivalent if their implied observed-data distributions are the same.
\end{definition}

Consider now a class of full-data distributions $\mathcal{C}_\Theta$ indexed by the parameter space $\Theta$ which is either finite- or infinite-dimensional.  If we were able to observe the values of the study variables regardless of the values of their missingness indicators, we would still have to guarantee that $\mathcal{C}_\Theta$ is identifiable in the usual sense \citep[e.g.,][p. 24]{LehmannCasella98}.  

\begin{definition}[Full-data identifiability] 
A class of full-data distributions $\mathcal{C}_\Theta$ is said to be full-data identifiable if there exists a bijection from $\Theta$ to $\mathcal{C}_\Theta$.
\end{definition}

Full-data identifiability is an elementary requirement of $\mathcal{C}_\Theta$ which simply says that the class is properly parameterized, and therefore throughout this article we will assume that this holds.  Let now $\text{obs}(\mathcal{C}_\Theta)$ denote the class of observed-data distributions implied by $\mathcal{C}_\Theta$ according to \eqref{eq:comp_obsdata}.  

\begin{definition}[Identifiability] A class of full-data distributions $\mathcal{C}_\Theta$ is said to be identifiable if there exist bijections from $\Theta$ to $\mathcal{C}_\Theta$ and from $\text{obs}(\mathcal{C}_\Theta)$ to $\mathcal{C}_\Theta$.
\end{definition}

The first bijection in this definition corresponds to full-data identifiability for $\mathcal{C}_\Theta$, and the second one simply tells us that we need a unique way to go back and forth from $\text{obs}(\mathcal{C}_\Theta)$ to $\mathcal{C}_\Theta$.  These two bijections imply a third one between $\text{obs}(\mathcal{C}_\Theta)$  and $\Theta$, which corresponds to the common notion of identifiability applied to $\text{obs}(\mathcal{C}_\Theta)$.

If the class of observed-data distributions $\text{obs}(\mathcal{C}_\Theta)$ is a proper subset of all observed-data distributions, then the model $\mathcal{C}_\Theta$ imposes parametric restrictions on what could be nonparametrically recovered from observed data alone.  Thus, we also make use of a  stricter property for a class of full-data distributions, namely, nonparametric identification, also known as nonparametric saturation or just-identification \citep{Robins97,Vansteelandtetal06, DanielsHogan08, HoonhoutRidder18}.  

\begin{definition}[Nonparametric identifiability] A class of full-data distributions $\mathcal{C}_\Theta$  is said to be nonparametrically identifiable if it is identifiable and $\text{obs}(\mathcal{C}_\Theta)$ equals the set of all observed-data distributions.
\end{definition}

Given the bijective mapping between $\Theta$ and $\text{obs}(\mathcal{C}_\Theta)$ obtained from the  identifiability requirement, we can think of a nonparametrically identifiable class as being indexed by the set of all observed-data distributions.  

Two nonparametrically identifiable classes necessarily lead to full-data distributions that are observationally equivalent, and therefore the assumptions used to build such classes cannot be refuted from observed data alone.  When two missing data models are observationally equivalent, any discrepancies in inferences are due entirely to the difference in the restrictions on the unidentifiable parts of the full-data distribution.  Nonparametric identification additionally guarantees that these restrictions do not constrain the observed-data distribution.  Nonparametric identification  is therefore a basic  desirable property, particularly useful for comparing inferences
under different  missing data assumptions.

\subsection{Auxiliary marginal information}\label{ss:ami}

To guarantee the nonparametric identifiability of the sequentially additive nonignorable missingness mechanisms that we will introduce, we need to have access to  
external information on features of the distribution of some of the study variables, here represented by $Y=(Y_1,\dots,Y_p)$.  Suitable examples include the joint distribution of $Y$, joint marginal distributions for subsets of $Y$, the individual marginal distributions of each $Y_j$, or expected values of functions of the $Y$ variables, including means, variances, and general moments.  We provide a formal definition that encompasses all of these cases.

\begin{definition}[Auxiliary marginal information]\label{def:ami} 
Let the set of functions $\mathcal{U}\subseteq L_1\{f(y)\}$ contain for each $j=1,\dots,p$ at least one almost-surely non-constant function of $y_j$. If we know the value of $E[u(Y)]=\int u(y) f(y)\mu(dy)<\infty$ for each $u\in\mathcal{U}$, we refer to $\{E[u(Y)]\}_{u\in\mathcal{U}}$ as auxiliary marginal information on the distribution of $Y=(Y_1,\ldots,Y_p)$.  
\end{definition}
 
For all of our results, to separate identification from estimation issues, having auxiliary marginal information on a set of variables is conceptualized as knowing the expected value of a set of functions $\mathcal{U}$ with respect to the distribution of $Y$.  Data-based implementations are discussed in Section \ref{s:imp}.  

Definition \ref{def:ami} includes different types of external information.  Important cases include the marginal distribution of $Y$, where $\mathcal{U}$ would be taken as all the integrable functions with respect to the distribution of $Y$; the marginal distributions of each $Y_j$, where $\mathcal{U}$ would be taken as all the integrable functions with respect to the distributions of each $Y_j$ seen as functions of the whole vector $y$; the means of the variables, where $\mathcal{U}$ would contain $p$ functions $u_1,\dots,u_p$ where $u_j(y)=y_j$; general moments of the variables, where $\mathcal{U}$ would contain functions $u(y)=y_j^z$ for $z\neq 0$; among many others.  Intuitively, the richer the set $\mathcal{U}$ the more flexible the class of missingness mechanisms that we will be able to identify. 

Auxiliary data on marginal distributions can be available in many contexts and in many forms.  For example, published data products from national censuses of  businesses, farms, and people provide marginal distributions for many variables commonly included in sample surveys of those populations, such as number of employees and demographic characteristics. Indeed, sample surveys routinely use known margins from censuses and sampling frames in weight calibration and generalized regression estimation \citep[e.g.,][ch. 11]{lohr10}.   Administrative databases, such as tax records, voter registration files, state education records on school children, and medical records from insurers or government programs, like Medicare in the U.S.A., can provide margins on potentially relevant populations.  Large nationally representative surveys like the American Community Survey in the U.S.A. can provide estimates of marginal means and totals with small enough standard errors to be treated as essentially known.  Marginal distributions of biomarker measurements may be available from calibration studies done by assay producers or government agencies.  Other examples of data analyses that use auxiliary information can be found in, for example, \cite{berrocal:mlm:gelfand}, \cite{Chatterjee16}, \cite{guo:little:mcconnell}, and \cite{nascombining}.

Naturally, analysts should carefully consider their choice of external data sources.  In particular, analysts should ensure that the information is contemporaneous with, covers the same target population as, and uses the same variable definitions as the data being analyzed. Analysts should feel comfortable that the information does not contain substantial biases, for example, from convenience sampling, measurement error, and nonresponse bias within the data used to obtain the auxiliary marginal information.

\section{Sequential additive nonignorability}\label{s:SAN}

\subsection{Extending results for univariate nonresponse}\label{univariate}

We begin with a single random variable $Y$ subject to missingness; here, $M$ denotes its missingness indicator, and the vector of variables $X$ is fully observed.  This is the set-up studied by 
\cite{Hiranoetal98,Hiranoetal01} in the context of a longitudinal study, where the $X$ variables are recorded at a given time point, and $Y$ denotes a follow-up measurement that is sometimes missing due to attrition. In that context the auxiliary marginal information comes from a refreshment sample, seen as a random sample from the marginal distribution of $Y$, which they conceptualize as knowing $f(y)$.

Given $f(y)$, \cite{Hiranoetal98,Hiranoetal01} showed that there exist identifiable missingness mechanisms with the form 
\begin{align}\label{eq:ANI}
\lambda[f(M=1\mid x, y)] & = \alpha(x)+\beta(y),
\end{align}
for a link function $\lambda$, and for some functions $\alpha$ and $\beta$ which are essentially unrestricted except for some integrability conditions.   The model is additive in $\alpha(x)$ and $\beta(y)$ because there is not enough information to identify interactions between $X$ and $Y$ \citep{Hiranoetal98,Hiranoetal01}.  Examples of analyses that make use of additive nonignorable missingness mechanisms include the work in \cite{Nevo03}, \cite{Bhattacharya08}, and \cite{SiReiterHillygus15}, among others.

The missingness mechanism in \eqref{eq:ANI} is appealing, as it includes as special cases a missing always at random mechanism \citep{MealliRubin15} when $\beta(y) = 0$ and the often-used selection model of \cite{hausman1979attrition} when $\alpha(x)=0$.  Thus, the additive nonignorable model can be viewed as an alternative to imposing one of those two missingness mechanisms, and instead letting the data determine a compromise.  

The existence results of \cite{Hiranoetal98,Hiranoetal01} are limited to distributions that are absolutely continuous with respect to the product Lebesgue measure \citep{Hiranoetal01} or that have finite support \citep{Hiranoetal98}, meaning that they do not cover problems with variables of mixed type.  Our goal in this section is to extend their identification results to general variable types, and to permit the usage of auxiliary marginal information that can be coarser compared to $f(y)$, as in Definition \ref{def:ami}, in which case $\beta$ in \eqref{eq:ANI} will be restricted to be in $\langle\mathcal{U}\rangle$.

To clearly separate identification from estimation issues, in our identification results we assume the observed-data density $f(x,y^*)$ to be known and availability of perfect auxiliary marginal information, corresponding to knowing the value of $E[u(Y)]$ for all functions $u\in\mathcal{U}$ as in Definition \ref{def:ami}.  Using a pattern mixture model formulation for missing data, the observed-data density $f(x,y^*)$ in this case corresponds to $\pi f(x\mid M=1)$ when $M=1$, and $(1-\pi)f(x,y\mid M=0)$ when $M=0$, where $\pi$ is the probability of $M=1$.   For each $u\in \mathcal{U}$ we also can derive each $E[u(Y)\mid M=0]=\int u(y) f(y\mid M=0)\mu(dy)$ using $f(y\mid M=0)=\int f(x,y\mid M=0) \nu(dx)$ from the observed-data distribution.  Combining these with auxiliary marginal information, we can obtain $E[u(Y)\mid M=1]=\int u(y) f(y\mid M=1)\mu(dy)=\{E[u(Y)]-(1-\pi)E[u(Y)\mid M=0]\}/\pi$.
This means that auxiliary marginal information allows us to find the value of integrals 
which are computed with respect to the distribution of $Y\mid M=1$ that cannot be obtained from the observed-data distribution.  Therefore, while $f(x,y\mid M=1)$ is unknown, we know its marginal $f(x\mid M=1)$ and have a set of constraints given by the values of the integrals $\int u(y) f(y\mid M=1)\mu(dy)$ for $u\in \mathcal{U}$.  

From an information theoretic point of view, it is natural to think of approximating the true $f(x,y\mid M=1)$ by an  information projection of $f(x,y\mid M=0)$ onto the set of distributions that have the $X$-marginal given by $f(x\mid M=1)$ and that also satisfy the constraints imposed by the auxiliary marginal information.  The $\fdiv$-divergence $I_{\fdiv}$ between distributions with densities $g^*(x,y)$ and $g(x,y)$ is given by 
$$I_{\fdiv}(g^*,g)=\iint \fdiv \left[\frac{g^*(x,y)}{g(x,y)}\right]g(x,y)\nu(dx)\mu(dy),$$
for a convex and differentiable function $\fdiv:(0,\infty)\mapsto \mathbb{R}$ \citep[][]{csiszar63}. For example, when $\fdiv(z)=z\log (z)$ we obtain the Kullback--Leibler divergence.  
The $\fdiv$-projection of a probability distribution with density $g(x,y)$ onto a set of probability distributions $\mathcal{C}$ is defined as the element in $\mathcal{C}$ with density $g^*(x,y)$ that minimizes $I_{\fdiv}(g^*,g)$ \citep[see, e.g.,][Ch. 8]{LieseVajda87}.   We will show that there is an intrinsic connection between the function $\fdiv$ used in $\fdiv$-projections and the link function $\lambda$ used to define additive nonignorable missingness mechanisms.  Our results rely on those of \cite{LieseVajda87}, which require $\fdiv$ not to increase too fast as its argument approaches infinity. 

\begin{assumption}[Regular link function]
\label{link}
The link function $\lambda:(0,1)\mapsto \mathbb{R}$ is differentiable and monotonically increasing.
\end{assumption}

\begin{assumption}[Growth of $\fdiv$]
\label{f-growth}
The function $\fdiv$ satisfies the property that for every $t>1$ there exist positive constants $t_0, t_1, t_2, t_3$ such that for all $z>t_0$, $\fdiv(tz)\leq t_1\fdiv(z)+t_2z+t_3$.
\end{assumption}

\begin{theorem}[Identification]\label{th:1} Let $X$ be a vector of always observed random variables, $Y$ be a random variable subject to missingness and $M$ be its missingness indicator.   Assume that the observed-data density $f(x,y^*)$ and auxiliary marginal information $\{E[u(Y)]\}_{u\in \mathcal{U}}$ are derived from a distribution that satisfies $\lambda[f(M=1\mid x, y)] = \alpha(x)+\beta(y)$, where $\lambda$ satisfies Assumption \ref{link}, $\alpha \in L_1\{f(x\mid M=1)\}$, and $\beta\in\langle\mathcal{U}\rangle$.  
Assume that $\fdiv_\lambda(z) = \int_{0}^z\lambda[v/(c+v)]dv$, $c=(1-\pi)/\pi$, satisfies Assumption \ref{f-growth}.  
Then $f(x,y\mid M=1)$ is the $\fdiv_\lambda$-projection of $f(x,y\mid M=0)$ onto the set of distributions that match both the marginal defined by $f(x\mid M=1)$ and the expectations given by $\{E[u(Y)\mid M=1]\}_{u\in \mathcal{U}}$.
\end{theorem}

All of our proofs are presented in Appendix 2.  They rely on some results on $\fdiv$-projections presented in Appendix 1.  Theorem \ref{th:1} indicates that under additive nonignorability, if the missingness mechanism satisfies $\lambda[f(M=1\mid x, y)] = \alpha(x)+\beta(y)$, with $\alpha \in L_1\{f(x\mid M=1)\}$ and $\beta\in\langle\mathcal{U}\rangle$, then the full-data distribution can be obtained from its implied observed-data distribution and from the auxiliary marginal information represented by $\{E[u(Y)]\}_{u\in \mathcal{U}}$, since we only need these pieces to recover $f(x,y\mid M=1)$ and thereby $f(x,y,m)$.  We are of course assuming that $\alpha$ and $\beta$ are  adequately set-up to ensure full-data identifiability of the class containing $f(x,y,m)$, as mentioned in Section \ref{identifiable}.  Theorem \ref{th:1} also requires $\fdiv_\lambda$ to satisfy Assumption \ref{f-growth}, for which a sufficient condition is that $\lim_{z\rightarrow \infty}\fdiv(z)/z^{a}=0$ for some $a>0$ \citep[][p. 171]{LieseVajda87}.  This is the case for common link functions, such as the logit, probit, complementary log-log, and the link functions proposed by \cite{Aranda-Ordaz81}, just to name some, for all of which $\lim_{z\rightarrow \infty}\fdiv_\lambda(z)/z^2=0$.

Theorem \ref{th:1} indicates the largest additive nonignorable missingness mechanism that we can identify given the available auxiliary marginal information.  If $E(Y)$ is all we have access to, this result says that the model $\lambda[f(M=1\mid x, y)] = \alpha(x)+by$, with $b\in \mathbb{R}$, is identifiable; that is, the missingness mechanism can include a main linear effect of $Y$, such as in the example 1.9 of \cite{LittleRubin02}.  If $\mathcal{U}=\{u_1,\dots,u_k\}$, then the model $\lambda[f(M=1\mid x, y)] = \alpha(x)+\sum_{j=1}^k b_ju_j(y)$ is identifiable.  If we know the marginal distribution of $Y$, then $\mathcal{U}$ can be taken as the set of all integrable functions, and this result says that the model $\lambda[f(M=1\mid x, y)] = \alpha(x)+\beta(y)$, with $\beta \in L_1\{f(y\mid M=1)\}$, is identifiable, which corresponds to the additive nonignorable mechanism of \cite{Hiranoetal98,Hiranoetal01}.

 The statement of Theorem \ref{th:1} suggests a plug-in approach for obtaining an estimate of $f(x,y\mid M=1)$, by computing the $\fdiv$-projection of an estimate of $f(x,y\mid M=0)$ onto the set of distributions that match estimates of $f(x\mid M=1)$ and $\{E[u(Y)\mid M=1]\}_{u\in \mathcal{U}}$.  While there exist algorithms for doing so \citep[e.g.,][]{Ruschendorf95,Bhattacharya06}, they can be challenging to implement as they require iterative approximation of potentially complex integrals \citep{Ruschendorf95}.  Instead, we rely on the theory of $\fdiv$-projections merely for  identification results, and present a more straightforward, likelihood-based implementation in Section \ref{s:imp}.

The following result indicates that a full-data distribution derived assuming additive nonignorability is observationally equivalent to the true full-data distribution.

\begin{theorem}[Nonparametric identification]\label{th:NPS_AN}
 Let the observed-data density $h(x,y^*)$  and the auxiliary marginal information $\{\int u(y) h(y)\mu(dy)\}_{u\in\mathcal{U}}$ be derived from a full-data density $h(x,y,m)$.  Let $\fdiv_\lambda(z) = \int_{0}^z\lambda[v/(c+v)]dv$, with $c=(1-\pi)/\pi$ and $\pi=h(M=1)$, satisfy Assumption \ref{f-growth}, for  a function $\lambda$ satisfying Assumption \ref{link}.  Let $g(x,y\mid M=1)$ denote the $\fdiv_\lambda$-projection of $h(x,y\mid M=0)$ onto the set of distributions that match the marginal defined by $h(x\mid M=1)$ and the integrals $\{\int u(y) h(y\mid M=1)\mu(dy)\}_{u\in\mathcal{U}}$.  Define a full-data distribution as $g(x,y,m)=\{g(x,y\mid M=1)\pi\}^m\{h(x,y,M=0)\}^{1-m}$.  Then $g(x,y,m)$ encodes an additive nonignorable missingness mechanism with 
$\lambda[g(M=1\mid x, y)] \in \overbar{\langle L_1\{h(x\mid M=1)\}\cup \mathcal{U}\rangle}$, and furthermore $g(x,y^*)=h(x,y^*)$ and $\int u(y) g(y)\mu(dy)=\int u(y) h(y)\mu(dy)$ for all $u\in\mathcal{U}$.
\end{theorem}
 This result indicates that if one derives a full-data density $g(x,y,m)$ assuming additive nonignorability from a given observed-data density $h(x,y^*)$  and  auxiliary marginal information $\{\int u(y) h(y)\mu(dy)\}_{u\in\mathcal{U}}$, then $g(x,y,m)$ implies back the original $h(x,y^*)$ and $\{\int u(y) h(y)\mu(dy)\}_{u\in\mathcal{U}}$.  In other words, additive nonignorability induces a one-to-one mapping from the set of observed-data distributions and auxiliary marginal information to the set of full-data distributions. A technical detail in Theorem \ref{th:NPS_AN} is that $\lambda[g(M=1\mid x, y)]$ need not be in $\langle L_1\{h(x\mid M=1)\}\cup \mathcal{U}\rangle$ but it could be a limit point outside of this set, although in that case $\lambda[g(M=1\mid x, y)]$ can be  arbitrarily approximated by functions of the form $\alpha(x)+\beta(y)$ where $\alpha \in L_1\{h(x\mid M=1)\}$, and $\beta\in\langle\mathcal{U}\rangle$.

\subsection{Multivariate nonresponse}

We now extend the concept of additive nonignorability to the context of multivariate item nonresponse, where each of the variables in $Y=(Y_1,\ldots,Y_p)$ is subject to nonresponse, with $M=(M_1,\ldots,M_p)$ being its vector of missingness indicators.  For now, we still consider the vector of variables $X$ to be fully observed, but we relax this requirement in Section \ref{ss:part_ign}. 

We begin by defining a comprehensive class of sequentially additive nonignorable missingness mechanisms that allows $M_j$ to depend directly on $Y_j$ for each $j$, yet also meets the criterion of nonparametric identification.  In some contexts, however, analysts may find it convenient to use submodels of the comprehensive version that we  introduce.  These may be easier to interpret or estimate, as we discuss in Section \ref{ss:special} and Section \ref{s:imp}.  The identification results for the comprehensive  version, however, provide assurance that its submodels also are identifiable, albeit without the advantages endowed by nonparametric identification.

We factorize the missingness mechanism as
$$f(m_1,\ldots,m_p\mid x,y)=\prod_{j=1}^p f(m_j\mid x,y, m_{<j}),$$
where we let $f(m_j\mid x,y, m_{<j})$  be as general as possible to obtain nonparametric identification.  
A similar sequential factorization strategy is used by \cite{ibrahim}, among others.  It requires us to impose an ordering on the $p$ variables.  To facilitate explanations, we   
proceed as if the variables $Y_1,\ldots,Y_p$ are indexed by the order in which they are collected, which is a natural choice in longitudinal studies or when we know the order in which questions are administered in a survey.  Of course, sequential additive nonignorability can be defined for any other ordering, and different orderings will lead to different missingness mechanisms.  We discuss guidelines for selecting orderings in Section \ref{ss:issues}. The order of the $X$ variables with respect to those in $Y$ is irrelevant.

To motivate the comprehensive version of the sequentially additive nonignorable missingness mechanism, we shall think of a hypothetical respondent from whom we attempt to collect values of the variables $Y_1,\ldots,Y_p$.  We start by trying to collect her value of $Y_1$, but she may or may not report it.  Whether she reports it or not is determined by a probabilistic mechanism $f(m_1 \mid x,y)$, which we assume to satisfy  \begin{align*}
\lambda\{f(M_1=1 \mid x,y)\}
& =  \alpha_1(x,y_{>1})+\beta_1(y),
\end{align*}
for some functions $\alpha_1$ and $\beta_1$ subject to constraints described later.  This  indicates that, given a value of $Y_{>1}$, the nonresponse for $Y_1$ follows an additive nonignorable mechanism as in \eqref{eq:ANI}.  The functions $\alpha_1(x,y_{>1})$ and $\beta_1(y)\equiv \beta_1(y_1,y_{>1})$   represent interactions between $X$ and $Y_{>1}$, and $Y_1$ and $Y_{>1}$, respectively, but the model does not allow interactions between $Y_1$ and $X$.  In particular, this means that the nonresponse for $Y_1$ can depend on $Y_1$, and this dependence can change across the values of $Y_{>1}$ but is homogeneous across the values of $X$.  The result of our attempt to measure $Y_1$ is a realization of its materialized variable $Y_1^*$.

We then attempt to measure the respondent's value of $Y_2$.  Whether she reports this value is determined by a probabilistic mechanism that we assume to satisfy $f(m_2 \mid x,y,m_{1}) =  f(m_2 \mid x,y^*_{1},y_{\geq 2})$, that is, the probability of nonresponse for $Y_2$ depends on $Y_1$ and $M_1$ only through the materialized variable $Y_1^*$, namely, if the value of $Y_1$ is not revealed then it does not influence the probability of nonresponse for $Y_2$.  
Since $Y_1^*$ captures all the information of $M_1$,  the probability of nonresponse for $Y_2$ does depend on whether $Y_1$ is reported.  We further assume that 
\begin{align*}
\lambda\{f(M_2=1 \mid x,y^*_{1},y_{\geq 2})\} 
& =  \alpha_2(x,y^*_{1},y_{>2})+\beta_2(y_{\geq 2}),
\end{align*}
for some functions $\alpha_2$ and $\beta_2$ described later.  Similarly as for the first item, for each value of $Y_{>2}$ the nonresponse for $Y_2$ follows an additive nonignorable mechanism, where $\alpha_2(x,y^*_{1},y_{>2})$ represents interactions between $(X,Y_1^*)$ and $Y_{>2}$, or equivalently, interactions between $(X,Y_1,M_1)$ and $Y_{>2}$ which are homogeneous across the missing values of $Y_1$.  The direct dependence of $M_2$ on $Y_2$ is captured by $\beta_2(y_{\geq 2})\equiv \beta_2(y_2,y_{>2})$, which allows this dependence to vary with $Y_{>2}$.  The dependence of $M_2$ on $Y_2$, however, is homogeneous across the values of  $(X,Y_1,M_1)$.  Thus far, the result of our data collection process is $(Y^*_1,Y^*_2)$. 

After having attempted to collect the respondent's values for the first $j-1$ variables, $Y_{<j}$, we have actually obtained a realization of their materialized variables $Y_{<j}^*$.  At this point, the missingness mechanism for whether we observe the respondent's value of $Y_j$ is defined by $f(m_j \mid x,y,m_{<j}) =  f(m_j \mid x,y^*_{<j},y_{\geq j})$.  The assumption in this mechanism is that the nonresponse for $Y_j$ does not depend on the missing values of the previous variables in the sequence, that is, its dependence on $Y_{<j}$ and $M_{<j}$ comes only through the materialized variables $Y^*_{<j}$.  We further assume that, for each value of $Y_{>j}$ the nonresponse mechanism for $Y_j$ is additive nonignorable, that is,
\begin{align*}
\lambda\{f(M_j=1 \mid x,y^*_{<j},y_{\geq j})\}
& = \alpha_j(x,y^*_{<j},y_{>j})+\beta_j(y_{\geq j}).
\end{align*}
Here, the function $\alpha_j(x,y^*_{<j},y_{>j})$ represents interactions between $(X,Y_{<j},M_{<j})$ and $Y_{>j}$, although these interactions are constant across the missing values of $Y_{<j}$.  Also, the function $\beta_j(y_{\geq j})\equiv \beta_j(y_j,y_{>j})$ represents interactions between $Y_j$ and $Y_{>j}$, but the model does not have an interaction between $Y_j$ and $(X,Y_{<j},M_{<j})$.  

The final step in our data collection attempt is to try to record the respondent's value of $Y_p$, at which point we have collected her value of $Y_{<p}^*$.  For the final variable, we assume the missingness mechanism to be $f(m_p \mid x,y,m_{<p}) =  f(m_p \mid x,y^*_{<p},y_{p})$, with
\begin{align*}
\lambda\{f(M_p=1 \mid x,y^*_{<p},y_{p})\}
& =  \alpha_p(x,y^*_{<p})+\beta_p(y_{p}),
\end{align*}
meaning that the nonresponse for the last variable does not depend on any of the missing values for the previous variables, but it can depend on the value of $Y_p$ itself, although this dependence is homogeneous across all the values of $(X,Y_{<p},M_{<p})$.

\begin{definition}[Sequential additive nonignorability]\label{def:SAN}
Let $X$ be a vector of always observed random variables, $Y$ be a vector of $p$ random variables subject to missingness and $M$ be its vector of missingness indicators.  A missingness mechanism is sequentially additive nonignorable if it can be written as
$f(m\mid x,y)  = \prod_{j=1}^p f(m_j \mid x,y,m_{<j})$, where for $j=1,\ldots,p$,
\begin{align}\label{eq:sani}
f(m_j \mid x,y,m_{<j})
& =  f(m_j \mid x,y^*_{<j},y_{\geq j}),
\end{align}
with
\begin{align}\label{eq:logitsani}
\lambda\{f(M_j=1 \mid x,y^*_{<j},y_{\geq j})\} & = \alpha_j(x,y^*_{<j},y_{>j})+\beta_j(y_{\geq j}),
\end{align}
where $\lambda$ is a link function, and $\alpha_j$ and $\beta_j$ are real-valued functions.
\end{definition}

The $\alpha_j$ and $\beta_j$ functions in this definition require some constraints to guarantee full-data identifiability.  In this article we constrain $\beta_j(y_j^0,y_{>j})=0$ for some arbitrary value $y_j^0$ of $Y_j$ and for all values $y_{>j}$ of $Y_{>j}$, while leaving the $\alpha_j$ functions unconstrained; this restriction implies that $\beta_j(y_{\geq j})$ cannot be expressed with additive terms that only depend on $y_{>j}$.  Other restrictions are possible; for example, one could add an intercept to the linear part in \eqref{eq:logitsani} while also constraining $\alpha_j$. 
In Theorems \ref{th:SAN_ident} and \ref{th:SAN_NPS} we impose further restrictions on the $\alpha_j$ and $\beta_j$ functions to guarantee identifiability and nonparametric identifiability.  The restrictions for $\beta_j$ are determined by the auxiliary marginal information on the distribution of $Y_{\geq j}$, specifically we require $\beta_j\in \langle \mathcal{U}_{\geq j}\rangle$, where $\mathcal{U}_{\geq j}$ denotes the set of functions in $\mathcal{U}$ that depend exclusively on $y_{\geq j}$.

Our proofs of identifiability and nonparametric identifiability rely on Algorithm \ref{al1}, which uses a sequence of information projections to construct a full-data distribution that satisfies sequential additive nonignorability, taking the observed-data density $f(x,y^*)$ and the auxiliary marginal information $\{E[u(Y)]\}_{u\in\mathcal{U}}$ as input.  The true full-data density, from which $f(x,y^*)$ and $\{E[u(Y)]\}_{u\in\mathcal{U}}$ are derived, is denoted $f(x,y,m)$, and the densities obtained from Algorithm \ref{al1} are denoted with $g$.  Theorem \ref{th:SAN_ident} shows identifiability, since if $f(x,y,m)$ truly satisfies sequential additive nonignorability, then the output $g(x,y,m)$ of Algorithm \ref{al1}  equals $f(x,y,m)$ almost surely.  Theorem \ref{th:SAN_NPS} shows nonparametric identifiability, since sequential additive nonignorability cannot be refuted using  $f(x,y^*)$ and $\{E[u(Y)]\}_{u\in\mathcal{U}}$ alone, given that $g(x,y,m)$ is observationally equivalent to the true $f(x,y,m)$.   

\begin{algorithm} Full-data distribution construction algorithm. \label{al1}
\begin{tabbing}
\qquad \enspace Input $g(x,y^*)\equiv f(x,y^*)$, $\{E[u(Y)]\}_{u\in\mathcal{U}}$.\\
\qquad \enspace For $j=p,\ldots,1$\\
\qquad \qquad a. Use $g(x,y^*_{\leq j},y_{> j})$ to derive $g(x,y^*_{<j},y_{\geq j}\mid M_j=0)$, $g(x,y^*_{<j},y_{> j}\mid M_j=1)$, \\
\qquad \qquad \qquad and $g(y_{\geq j},M_j=0)$, and $\pi_j=g(M_j=1)$.\\
\qquad \qquad b. For each $u\in \mathcal{U}_{\geq j}$ compute \\ 
\qquad \qquad \qquad $E_g[u(Y_{\geq j})\mid M_j=1]=\{E[u(Y_{\geq j})]-\int u(y_{\geq j})g(y_{\geq j},M_j=0)\mu(dy_{\geq j})\}/\pi_j$\\
\qquad \qquad c. Find $g(x,y^*_{<j},y_{\geq j}\mid M_j=1)$ 
as the $\fdiv_{\lambda,j}$-projection of $g(x,y^*_{<j},y_{\geq j}\mid M_j=0)$ \\
\qquad \qquad \qquad onto the set of distributions that match the marginal $g(x,y^*_{<j},y_{> j}\mid M_j=1)$\\
\qquad \qquad \qquad and the expectations $\{E_g[u(Y_{\geq j})\mid M_j=1]; u\in \mathcal{U}_{\geq j}\}$,\\
\qquad \qquad \qquad with $\fdiv_{\lambda,j}(z) = \int_{0}^z\lambda[v/(c_j+v)]dv$, $c_j=(1-\pi_j)/\pi_j$. \\
\qquad \qquad d. Obtain 
$g(x,y^*_{<j},y_{\geq j})=\sum_{m_j=0}^{1}g(x,y^*_{<j},y_{\geq j}\mid M_j=m_j)\pi_j^{m_j}(1-\pi_j)^{1-m_j},$\\
\qquad \qquad \qquad and 
$g(m_j\mid x,y^*_{<j},y_{\geq j})=\frac{g(x,y^*_{<j},y_{\geq j}\mid M_j=m_j)}{g(x,y^*_{<j},y_{\geq j})}\pi_j^{m_j}(1-\pi_j)^{1-m_j}.$\\
\qquad \enspace Output $g(x,y,m)= g(x,y)\prod_{j=1}^p g(m_j\mid x,y^*_{<j},y_{\geq j})$.
\end{tabbing}
\end{algorithm}

\begin{theorem}[Identification]\label{th:SAN_ident} Let $X$ be a vector of always observed random variables, $Y$ be a random vector subject to missingness and $M$ be its vector of missingness indicators.   Assume that the observed-data density $f(x,y^*)$ and auxiliary marginal information $\{E[u(Y)]\}_{u\in \mathcal{U}}$ are derived from a distribution with density $f(x,y,m)$ that encodes a sequentially additive nonignorable missingness mechanism as in Definition \ref{def:SAN},  where $\lambda$ satisfies Assumption \ref{link}, $\alpha_j \in L_1\{f(x,y^*_{<j},y_{>j}\mid M_j=1)\}$, and $\beta_j\in\langle\mathcal{U}_{\geq j}\rangle$ for all $j=1,\ldots,p$. 
Assume that each $\fdiv_{\lambda,j}(z) = \int_{0}^z\lambda[v/(c_j+v)]dv$, with $c_j=(1-\pi_j)/\pi_j$ and $\pi_j=f(M_j=1)$, satisfies Assumption \ref{f-growth}.  Then 
\begin{enumerate}
\item $f(x,y^*_{<j},y_{\geq j}\mid M_j=1)$ is the $\fdiv_{\lambda,j}$-projection of $f(x,y^*_{<j},y_{\geq j}\mid M_j=0)$ onto the set of distributions that match the marginal defined by $f(x,y^*_{<j},y_{>j}\mid M_j=1)$ and the expectations given by $\{E[u(Y_{\geq j})\mid M_j=1]; u\in \mathcal{U}_{\geq j}\}$, for all $j=1,\ldots,p$. 
\item The output $g(x,y,m)$ of Algorithm \ref{al1} equals $f(x,y,m)$ almost surely.
\end{enumerate}
\end{theorem}

\begin{theorem}[Nonparametric identification]\label{th:SAN_NPS}
Let the observed-data density $h(x,y^*)$  and the auxiliary marginal information $\{\int u(y) h(y)\mu(dy)\}_{u\in\mathcal{U}}$ be derived from a full-data density $h(x,y,m)$.  Let $\lambda$ satisfy Assumption \ref{link}.  Let $\fdiv_{\lambda,j}(z) = \int_{0}^z\lambda[v/(c_j+v)]dv$, with $c_j=(1-\pi_j)/\pi_j$ and $\pi_j=h(M_j=1)$, satisfy Assumption \ref{f-growth} for each $j=1,\dots,p$. Let $g(x,y,m)$ be constructed as in Algorithm \ref{al1}.  Then
\begin{enumerate}
\item $g(x,y,m)$ encodes a sequentially additive nonignorable missingness mechanism with 
$\lambda[g(M_j=1\mid x, y^*_{<j}, y_{\geq j})] \in \overbar{\langle L_1\{g(x, y^*_{<j}, y_{>j}\mid M_j=1)\}\cup \mathcal{U}_{\geq j}\rangle}$, $j=1,\dots,p$.
\item The output $g(x,y,m)$ of Algorithm \ref{al1} and $h(x,y,m)$ are observationally equivalent, that is $g(x,y^*)=h(x,y^*)$ almost surely, and $\int u(y) g(y)\mu(dy)=\int u(y) h(y)\mu(dy)$ for all $u\in\mathcal{U}$.
\end{enumerate}
\end{theorem}

Algorithm \ref{al1} suggests a plug-in implementation, starting from estimates of the observed-data distribution and the auxiliary marginal information.  However, as we previously mentioned, we only use the theory of information projections for our identification results, and provide likelihood-based implementations in Section \ref{s:imp}.  Therefore, we use Algorithm \ref{al1} only as a theoretical tool for guaranteeing identification and nonparametric identification under sequential additive nonignorability.

\subsection{Applying the identification results}\label{ss:special}

In some settings, it may be difficult to conceptualize a response process that fully corresponds to 
the comprehensive version of sequential additive nonignorability. 
However, as we illustrate in this section, the comprehensive version includes important special cases of missingness mechanisms that are readily amenable to interpretation.  The identification results in Section \ref{s:SAN} assure analysts that models using these special-case missingness mechanisms can be estimated with enough data plus auxiliary marginal information.  In contrast, models that violate the constraints in \eqref{eq:sani} or \eqref{eq:logitsani} may not be identifiable even with infinite amounts of data.  Thus, analysts can use the identification results to specify missingness mechanisms that are interpretable submodels and know that these are identifiable.  Furthermore, analysts can work with the comprehensive version or with a large subclass of sequentially additive nonignorable models to enable data-driven compromises between simpler cases.  

As with the univariate additive nonignorable mechanism, the multivariate mechanism encodes ignorable and nonignorable cases.  When we set $\beta_j(y_{\geq j}) = 0$  and $\alpha_j(x, y_{< j}^*, y_{>j}) = \alpha_j(x, y_{< j}^*)$ for all $j$, we have an ignorable missingness mechanism as it does not depend on missing values. 
When we set $\alpha_j(x, y_{< j}^*, y_{ \geq j}) = \alpha_j$ and $\beta_j(y_{\geq j}) = \beta_j(y_{j})$ for all $j$, that is, nonresponse  for $Y_j$ depends only on the value of $Y_j$ itself, we encode a multivariate version of the selection model of  \cite{hausman1979attrition}.  Combining these two cases, taking $\alpha_j(x, y_{< j}^*, y_{>j}) = \alpha_j(x, y_{< j}^*)$ and $\beta_j(y_{\geq j}) = \beta_j(y_j)$ for all $j$, corresponds to the attrition mechanism of \cite{HoonhoutRidder18} in the case of monotone nonresponse. This also encodes an identifiable nonignorable nonresponse process with nonmonotone nonresponse.  An example where this special case of the comprehensive mechanism could be plausible is when $Y_{1},\ldots,Y_p$ are a set of medical tests that are administered sequentially.  Here, $M_j=0$ indicates that  test $j$ was performed on the patient, and  $M_j=1$ indicates otherwise. For some tests, we may have marginal distributions of the outcomes, say from baseline studies or meta-analyses.   It is reasonable to assume that physicians might not administer test $j$ on some patient because she is extremely confident that the patient's $Y_j$ would be in a medically safe range.  However, the physician's decision to administer the test also could depend on demographic characteristics $X$ and on values of previously administered tests $Y_{<j}^*$.  It seems plausible that the potential results of tests that were not previously administered have no influence on the physician's decision to administer a new test.

A further simplification of the previous subclass can be obtained by seting $\alpha_j(x, y_{< j}^*, y_{>j}) = \alpha_j(x)$ and $\beta_j(y_{\geq j}) = \beta_j(y_j)$ for all $j$. This allows nonresponse for $Y_j$ to depend on its values, as well as the variables in $X$, while ensuring that inferences are invariant to the ordering of $Y_1,\ldots,Y_p$.  This can be convenient when analysts have no reasonable assumptions on which to base an ordering.  Fully observed variables in $Y$ can be put in any order with respect to those with missingness, given that we do not model their missingness mechanism.  When some of the $Y$ variables are fully observed, say without loss of generality $Y_{\geq j'}$, the submodel where $\alpha_j(x, y_{< j}^*, y_{>j}) = \alpha_j(x,y_{\geq j'})$ and $\beta_j(y_{\geq j}) = \beta_j(y_j, y_{\geq j'})$, for $j<j'$, is also invariant to the ordering of the variables.    An illustration of this situation appears in the example of Section \ref{s:appl}. 

Another feature of the general missingness mechanism is that nonresponse for a variable $Y_j$ can, although it does not have to, depend on the following variables in the sequence $Y_{>j}$.  For example, suppose a survey collects, among other items, $Y_{j}$ which indicates whether voted in the last election, and $Y_{j'}$ which indicates political affiliation; here, $j<j'$.  Marginal information about these variables is available from external sources.  It seems plausible that the probability of reporting voting turnout might depend on the political affiliation and on whether the respondent voted, regardless of whether these variables get reported.  An advantage of sequentially additive nonignorable mechanisms is that such associations can be picked up if they exist.

Finally, an interesting connection is obtained by reversing the order of the items $Y_1,\ldots,Y_p$ and fixing $\beta_j(y_{\geq j})=0$ for all $j$, leading to the permutation missingness mechanism of \cite{Robins97}. That mechanism says that given an ordering of the study variables, the nonresponse propensity for variable $j$ depends on the values of the previous study variables in the order, whether observed or not, but not on variable $j$ nor on the following missing values in the order, which is the reverse of the interpretation that we have given.  \cite{Robins97} emphasized, however, that a limitation of the permutation missingness mechanism is that it does not allow the probability of missingness for a particular variable to depend on the value of that variable.  With sequential additive nonignorability, we are not subject to this limitation as the auxiliary marginal information allows us to obtain $\beta_j(y_{\geq j})\neq 0$.  

As we can see, sequential additive nonignorability leads to very flexible classes of missingness models, as it encompasses a number of important particular cases that encode potentially plausible missingness mechanisms.

\section{Implementation}\label{s:imp}

\subsection{Practical considerations}\label{ss:issues}

The particular ordering of the variables encodes assumptions about the missingness mechanism and hence affects inferences.  Of course, orderings imply distributional assumptions for any multivariate missing data modeling strategy based on chained conditional distributions \citep{ibrahim, danielsbart}.  Depending on the context, some orderings may lead to assumptions deemed more plausible than others. 
In some contexts,  it may be reasonable to order variables temporally, for example, following the sequence in which questions are asked in a survey questionnaire or time points in a longitudinal study.  In other settings, natural orderings may not be apparent.  In this case, analysts can view the comprehensive version of sequential additive nonignorability as a rich class of multivariate nonignorable missing data models that allow $M_j$ to depend on $Y_j$, as well as allow estimation of additional dependencies that the analyst may not have considered initially.  In such cases, it may be computationally convenient to order variables from highest to lowest fractions of missing data, so that the richest models are used for the variables with the most missing data.   Alternatively, analysts could use the simpler mechanism described in Section \ref{ss:special} that does not require any ordering. When the ordering is somewhat arbitrary, it is prudent for analysts to analyze the sensitivity of results to different orderings, as we do in Section \ref{s:appl}.

Furthermore, in practice one works with finite samples, and therefore working with models with the most comprehensive structure in \eqref{eq:logitsani} may lead to complications.
For example, if some category of a categorical variable, or a combination of categories of different categorical variables, happens not to be observed in the sample, then their corresponding parameters are not estimable from the likelihood function alone, as it will be constant as a function of those parameters.  In such cases, maximum likelihood estimates will not be unique, and Bayesian inference reliant on Markov chain Monte Carlo will suffer from convergence issues, unless strongly informative priors are imposed on those parameters.  Similar issues would occur if the number of unique model parameters exceeds the sample size. Thus, in many practical circumstances it is prudent to work with a model that respects the form of \eqref{eq:logitsani} but that does not include all possible interactions within the variables in $(X,Y^*_{<j},Y_{>j})$ nor within the variables in $Y_{\geq j}$.

\subsection{Partially ignorable multivariate nonresponse}\label{ss:part_ign}

We now return to the general set-up introduced in Section \ref{s:setup}, where some or all of the variables in the vector $X$, for which we do not have auxiliary marginal information, may also be subject to missingness, with $W$ denoting its vector of missingness indicators. 
In such cases, assuming sequential additive nonignorability for the missingness in both $Y$ and $X$ would lead to nonidentifiable models,  since we only have auxiliary marginal information for $Y$.  Therefore, to implement sequential additive nonignorability in those situations, we assume partial ignorability of the missingness mechanism \citep{HarelSchafer09}.  Specifically, we can write the missingness mechanism as
\begin{align}\label{eq:pi_miss_mech}
f(w,m\mid x,y) & = f(m\mid x,y)f(w \mid x,y,m),
\end{align}
where we assume $f(m\mid x,y)$ to be sequentially additive nonignorable, and we assume the missingness of the $X$ variables to be partially missing always at random \citep{HarelSchafer09, MealliRubin15}, that is,
\begin{equation}\label{eq:pmar}
f(w \mid x,y,m)=f(w \mid x_{\bar w},y_{\bar m}, m)\equiv f(w \mid x_{\bar w},y^*),
\end{equation}
for all $w\in\{0,1\}^q$ and $m\in\{0,1\}^p$.  This assumption indicates that the probability of observing missingness pattern $w$ in the $X$ variables does not depend on the unobserved values in $X$ and $Y$.  We take \eqref{eq:pmar} as an assumption on the missingness mechanism and not only on a specific realized value $w$.  
For related discussions see \cite{Seamanetal13} and \cite{MealliRubin15}.

The assumption in \eqref{eq:pmar} can be decomposed into two parts.  First, $f(w \mid x,y,m)=f(w \mid x,y^*)$, which says that this missingness mechanism is homogeneous across the missing values of the $Y$ variables.  In such case, \cite{SadinleReiter18} showed that if the missingness mechanism $f(w \mid x,y^*)$ leads to  nonparametric identified $f(x,w\mid y^*)$ for each $y^*$, and if $f(m\mid x,y)$ leads to nonparametric identified $f(x,y,m)$, then the combined missingness mechanism $f(m\mid x,y)f(w \mid x,y^*)$ leads to a nonparametric identified full-data distribution $f(x,y,w,m)$.  
The second part of the assumption in \eqref{eq:pmar} is $f(w \mid x,y^*)=f(w \mid x_{\bar w},y^*)$, which is a missing always at random assumption conditional on $Y^*$.  \cite{Gilletal97} showed that the missing always at random assumption leads to nonparametric identified distributions.
Therefore, following \cite{SadinleReiter18}, we obtain the property of nonparametric identification for full-data distributions derived under \eqref{eq:pi_miss_mech} and \eqref{eq:pmar}, with $f(m\mid x,y)$ being sequentially additive nonignorable.

\subsection{Likelihood-based inference}\label{ss:likelihood}

We consider the scenario where our initial goal is to use a random sample $\{(x_i,y_i)\}_{i=1}^n$ from a distribution with density $f(x\mid y, \theta)f(y\mid \kappa)$ to draw inferences on parameter vectors $\theta$ and $\kappa$. 
Here, the auxiliary marginal information about the distribution of $Y$ is included via the parameters $\kappa$. 
 If the sample is subject to missingness, we instead think of a full-data random sample $\{(x_i,y_i,w_i,m_i)\}_{i=1}^n$ drawn from a full-data distribution with density 
\begin{align}\label{eq:ell}
f(w\mid x,y,m,\phi)f(m\mid x,y,\gamma)f(x\mid y, \theta)f(y\mid \kappa)\equiv \ell(\phi,\gamma,\theta,\kappa ; ~ x,y,w,m),
\end{align}
with $\phi$ and $\gamma$ parameterizing the missingness mechanism.  
The full-data likelihood function is therefore $L(\phi,\gamma,\theta,\kappa)=\prod_{i=1}^n \ell(\phi,\gamma,\theta,\kappa ; ~ x_i,y_i,w_i,m_i)$.
The full-data random sample gets materialized as an observed-data random sample $\{(x_{i,\bar w_i},y_{i,\bar m_i},w_i,m_i)\}_{i=1}^n\equiv \{(x^*_i,y^*_{i})\}_{i=1}^n$.  The observed-data likelihood function is derived by integrating $L$ over the missing values $y_{i,m_i}$ and $x_{i,w_i}$, according to each missingness pattern $m_i$ and $w_i$, that is, 
$L_{obs}(\phi,\gamma,\theta,\kappa)=\prod_{i=1}^n \ell_{obs}(\phi,\gamma,\theta,\kappa ; ~ x_{i,\bar w_i},y_{i,\bar m_i},w_i,m_i)$, 
with
\begin{align*}
\ell_{obs}(\phi,\gamma,\theta,\kappa ; ~ x_{\bar w},y_{\bar m},w,m)=\iint \ell(\phi,\gamma,\theta,\kappa ; ~ x,y,w,m) \mu(dy_{m})\nu(dx_{w}).
\end{align*}
It is easy to check that taking $f(w\mid x,y,m,\phi)$ in \eqref{eq:ell} to be partially missing always at random, as in \eqref{eq:pmar}, implies that this part of the missingness mechanism can be ignored from the likelihood function for inference on $\gamma$, $\theta$ and $\kappa$.  We therefore work with the likelihood function 
$L_{obs}(\gamma,\theta,\kappa)=\prod_{i=1}^n \ell_{obs}(\gamma,\theta,\kappa ; ~ x_{i,\bar w_i},y_{i,\bar m_i},w_i,m_i)$, where
\begin{align*}
\ell_{obs}(\gamma,\theta,\kappa ; ~ x_{\bar w},y_{\bar m},w,m)=\iint f(m\mid x,y,\gamma)f(x\mid y, \theta)f(y\mid \kappa) \mu(dy_{m})\nu(dx_{w}).
\end{align*}
Taking $f(m\mid x, y,\gamma)$ as sequentially additive nonignorable permits us to write it as a product of logistic regressions $\prod_{j=1}^p f(m_j \mid x,y,m_{<j},\gamma_j)$, where 
\begin{align}\label{eq:logitj}
\lambda[f(M_j=1 \mid x,y,m_{<j},\gamma_j)]& =  \alpha_j(x,y^*_{<j},y_{>j})+\beta_j(y_{\geq j}),
\end{align}
with $\gamma_j$ representing the parameter functions $\alpha_j$ and $\beta_j$.  The nature of these functions depends on the type of variables in $X$ and $Y$, and on the auxiliary marginal information on the distribution of $Y$.  If we know $f(y)$, in the case of having only categorical variables in $X$ and $Y$, under the constraints $\beta_j(y_j^0,y_{>j})=0$ for arbitrary $y_j^0$ and for all $y_{>j}$, we have that the value of $\alpha_j$ for each possible value of $(X,Y^*_{<j},Y_{>j})$ and the value of $\beta_j$ for each possible value of $(Y_j,Y_{>j})$, $Y_j\neq y_j^0$, represent the parameters of the full model \eqref{eq:logitj}.  While this model is nonparametrically identified, in practice we may encounter issues estimating its parameters due to the finiteness of the sample, as we discuss in Section \ref{ss:issues}.  If $X$ and $Y$ contain continuous variables, the $\alpha_j$ and $\beta_j$ functions could be modeled using splines or Gaussian processes \citep[e.g.][]{Choudhurietal07},  which although not strictly nonparametric can be flexible enough to capture complex distributional features.  If the auxiliary marginal information is simply a finite set of moment restrictions $\{E[u(Y)]\}_{u\in \mathcal{U}}$ for $\mathcal{U}=\{u_1,\dots,u_k\}$, then it is easy to specify $\beta(y)=\sum_{j=1}^k b_ju_j(y)$.

Our final working likelihood is $L_{obs}(\gamma,\theta,\kappa)A(\kappa)$, where $A(\kappa)$ is a function whose form depends on the nature of the auxiliary marginal information.  If we have access to the true $\kappa$, for example if $Y$ is categorical and we know its true distribution from a census, then $A(\kappa)$ is simply an indicator function that equals zero when $\kappa$ is different from its census value.  If we have access to an additional fully observed random sample $\{y_i\}_{i=n+1}^m$ from the distribution of $Y$, as with refreshment samples, then $A(\kappa)=\prod_{i=n+1}^m f(y_i\mid \kappa)$.  If we have an estimate $\hat\kappa$ coming from a survey, then $A(\kappa)=f(\hat\kappa\mid \kappa)$ is the density function from the approximate distribution of $\hat\kappa$, such as the normal distribution in the case of Horvitz--Thompson estimators with large samples \citep[e.g.,][Chapter 2]{sarndal:swenson:wretman}.

\section{Illustrative example}\label{s:appl}

\subsection{Description of data and models}
Each year in the U.S.A., the Behavioral Risk Factor Surveillance System collects data on risk factors associated with a variety of diseases.  The data come from a random sample of adults contacted through a telephone survey \citep{BRFSS10}.  We use the 2010 data to estimate the prevalence of diabetes among demographic strata formed by combinations of age, race, and sex.  We focus on the U.S.\ Virgin Islands, as this territory has the highest nonresponse rates for 2010 in the variables that we study.  Our study variables include diabetes, defined as $X\in\{\textsc{no},\textsc{yes}\}$, with 0.17\% of nonresponse; age, defined as $Y_1\in\{\textsc{20--34},\textsc{35--49},\textsc{50--64},\textsc{65+}\}$, with 2.47\% of nonresponse; race, defined as $Y_2\in\{\textsc{black},\textsc{white},\textsc{other}\}$, with 5.89\% of nonresponse; and sex, defined as $Y_3\in\{\textsc{male},\textsc{female}\}$, being fully observed.  This ordering comes from the sequence in which the variables are recorded in the survey.  The joint distribution of age, race and sex in the U.S.\ Virgin Islands is available from the 2010 decennial census.  

We model $(X\mid Y_1=y_1,Y_2=y_2,Y_3=y_3) \sim \text{Bernoulli}\{\theta(y_1,y_2,y_3)\}$, with $\theta(y_1,y_2,y_3)$ representing the prevalence of diabetes in the stratum $(y_1,y_2,y_3)$, and we fix the marginal distribution of $(Y_1,Y_2,Y_3)$ at its census value.  We take a Bayesian approach to estimation and place flat priors on each $\theta(y_1,y_2,y_3)$.  The missingness mechanism for $X$ conditionally on $(X,Y,M)$ is assumed to be ignorable, as explained in Section \ref{ss:part_ign}.  We explore the estimation of the different per-stratum prevalences $\theta(y_1,y_2,y_3)$ under six different submodels of a full sequentially additive nonignorable missingness mechanism for $(M\mid X,Y)$, summarized in Table \ref{t:models}.

The first model that we consider is the full logit sequentially additive nonignorable missingness mechanism, which for the age nonresponse uses $\text{logit} f(M_1=1 \mid x,y)  =  \alpha_1(x,y_{2},y_3) + \beta_1(y_1,y_2,y_3)$, with the values of $\alpha_1$ for each $(x,y_{2},y_3)$ and $\beta_1$ for each $(y_1,y_2,y_3)$ being the parameters of the model.  This is equivalent to specifying the model in terms of a linear predictor using indicator variables for each $(x, y_2, y_3)$ and for each $(y_1, y_2, y_3)$.  With the constraint $\beta_1(Y_1=\textsc{20--34},y_2,y_3)=0$, we obtain $$\alpha_1(x,y_{2},y_3)= \text{logit} ~ f(M_1=1 \mid x,Y_1=\textsc{20--34},y_{2},y_3),$$ and for $y_1\neq \textsc{20--34}$,
$$\beta_1(y_1,y_2,y_3)=\log \frac{f(M_1=1 \mid x,Y_1=y_1,y_2,y_3)/f(M_1=0 \mid x,Y_1=y_1,y_2,y_3)}{f(M_1=1 \mid x,Y_1=\textsc{20--34},y_2,y_3)/f(M_1=0 \mid x,Y_1=\textsc{20--34},y_2,y_3)}.$$
This log-odds ratio explicitly captures the dependence of the nonresponse for the age variable on the categories of age.  This dependence is constant across the diabetes status, but it can vary with race and sex.  
For the race question, the full logit sequentially additive nonignorable missingness mechanism given $M_1=m_1$ uses 
\begin{align}
\text{logit} ~f(M_2=1 \mid x,y,m_1)  & = \text{logit} ~f(M_2=1 \mid x,y^*_{1},y_{\geq 2}) 
 = \alpha_2(x,y^*_{1},y_{3})+\beta_2(y_{2},y_3).\label{eq:m2}
\end{align}
With the constraint $\beta_2(\textsc{black},y_3)=0$, we have $$\alpha_2(x,y^*_{1},y_{3})=\text{logit} ~ f(M_2=1 \mid x,y_1^*,Y_{2}=\textsc{black},y_3),$$ and for $y_2\neq \textsc{black}$,
$$\beta_2(y_{2},y_3)=\log \frac{f(M_2=1 \mid x,y_1^*,Y_{2}=y_2,y_3)/f(M_2=0 \mid x,y_1^*,Y_{2}=y_2,y_3)}{f(M_2=1 \mid x,y_1^*,Y_{2}=\textsc{black},y_3)/f(M_2=0 \mid x,y_1^*,Y_{2}=\textsc{black},y_3)}.$$
This log-odds ratio measures the association between the race variable and its nonresponse, which might be different across sex.  
We use independent normals with mean zero and standard deviation 1.5 as priors for each $\alpha_1(x,y_{2},y_3)$ and each $\alpha_2(x,y^*_{1},y_{3})$, which lead to only slightly informative priors on the probability scale, and independent normals with mean zero and standard deviation 3 for each $\beta_1(y_1,y_2,y_3)$ and each $\beta_2(y_{2},y_3)$,  which are relatively spread out on the logit scale. 

We can give plausible interpretations to the components of this comprehensive model. In particular, we now describe a scenario where the propensity to respond to the race question $Y_2$ depends on the materialized variable for age $Y_1^*$, as assumed in \eqref{eq:m2}.  Holding $(X,Y_3)$ constant, consider two groups of people, those who do and do not respond to the age question. These groups could have different propensities to respond to the race question. For example, the first group could mostly include people who are willing to provide information to government agencies and hence are likely to respond to the race question, while the second group could mostly include people who believe that neither their age nor race---potentially sensitive variables---are the government's business and hence are unlikely to respond to the race question.  For this second group, the individuals may decide whether or not to respond to the race question independent of their actual age; for example, everyone in this group may be sufficiently distrustful or disinterested in the survey so as not to respond to the sensitive questions.
On the other hand, in the first group with generally response-compliant participants, it may be that younger people are less likely to report their race values than older people.  For example, the younger people may feel that the categories of race listed on the survey do not describe their actual race, making them less likely to answer the question. Or, these younger participants may be more likely than older participants to believe that race, but not age, is a private matter, and hence less likely to respond than older people.  
As discussed in Section \ref{ss:special}, the comprehensive version of the sequentially additive nonignorable missingness mechanism includes various ignorable and nonignorable models as special cases.  Thus, using the comprehensive version is appropriate even if a plausible missingness mechanism is actually represented by a submodel.  

In Table \ref{t:models} we summarize five submodels of the comprehensive version presented above.  The first submodel only has main effects, but still allows the nonresponse to directly depend on each item.  The second and third submodels are invariant to the ordering of the variables, with the third one representing a mechanism where the nonresponse directly depends only on the variable that we attempt to measure.  In the fourth submodel the nonresponse for each item does not directly depend on the item itself, but it is still nonignorable since $f(M_1=1 \mid x,y)$ depends on the unobserved $y_2$ values.  The fifth submodel corresponds to an ignorable mechanism.  The terms in each of these submodels have analogous interpretations as those in the full model, and therefore we similarly impose normal priors with mean zero and standard deviation 1.5 on each $\alpha$-term, and normals with mean zero and standard deviation 3 for the non-zero $\beta$-terms. 

We obtained approximate posterior distributions for the parameters of these six models using a standard Gibbs sampler, with a data augmentation scheme for the missing data \citep{TannerWong87}, and the strategy of \cite{PolsonScottWindle13} to expand the parts of the likelihood functions coming from the logistic regressions in terms of P\'olya--Gamma latent variables.  

\begin{table}
\def~{\hphantom{0}}
\caption{Subclasses of sequential additive nonignorability explored in Section \ref{s:appl}.}{%
\begin{tabular}{lcc}
\hline
 & $\text{logit} f(M_1=1 \mid x,y)$ & $\text{logit} f(M_2=1 \mid x,y^*_{1},y_{\geq 2})$ \\\hline
0. Full & $\alpha_1(x,y_{2},y_3) + \beta_1(y_1,y_2,y_3)$ & $\alpha_2(x,y^*_{1},y_{3})+\beta_2(y_{2},y_3)$ \\
\hspace{1cm} Restrictions &\footnotesize{$\beta_1(Y_1=\textsc{20--34},y_2,y_3)=0$}&\footnotesize{$\beta_2(Y_2=\textsc{black},y_3)=0$} \\[5pt]
1. Main effects & $\alpha_1(x)+\sum_{j=1}^3\beta_{1j}(y_j)$ & $\alpha_2(x)+\beta_{21}(y^*_{1})+\sum_{j=2}^3\beta_{2j}(y_{j})$\\
\hspace{1cm} Restrictions &\hspace{-.5cm}\footnotesize{$\beta_{11}(\textsc{20--34})=\beta_{12}(\textsc{black})=\beta_{13}(\textsc{male})=0$}&\footnotesize{$\beta_{21}(*)=\beta_{22}(\textsc{black})=\beta_{23}(\textsc{male})=0$}\\[5pt]
2. Order-invariant & $\alpha_1(x,y_3)+\beta_1(y_1,y_3)$ & $\alpha_2(x,y_3)+\beta_2(y_2,y_3)$\\
\hspace{1cm} Restrictions &\footnotesize{$\beta_1(Y_1=\textsc{20--34},y_3)=0$}&\footnotesize{$\beta_2(Y_2=\textsc{black},y_3)=0$} \\[5pt]
3. Only-directly dependent & $\alpha_1+\beta_1(y_1)$ & $\alpha_2+\beta_2(y_2)$\\
\hspace{1cm} Restrictions &\footnotesize{$\beta_{1}(Y_1=\textsc{20--34})=0$}&\footnotesize{$\beta_{2}(Y_2=\textsc{black})=0$}\\[5pt]
4. Not-directly dependent & $\alpha_1(x,y_{2},y_3)$ & $\alpha_2(x,y^*_{1},y_{3})$\\[5pt]
5. Ignorable & $\alpha_1(x)$ & $\alpha_2(x, y_{1}^*)$\\\hline
\end{tabular}}
\label{t:models}
\end{table}

\subsection{Results}
\begin{figure}[!t]
  \centering
  \begin{minipage}[b]{1\textwidth}
 \def\~{\hphantom{0}}
  \begin{tabular*}{\columnwidth}{c@{\extracolsep{\fill}}c@{\extracolsep{\fill}}c@{\extracolsep{\fill}}c@{\extracolsep{\fill}}c@{\extracolsep{\fill}}c@{\extracolsep{\fill}}}
   & \ \ & \multicolumn{4}{c}{Age} \\
   Race & & \textsc{20--34} & \textsc{35--49} & \textsc{50--64} & \textsc{65+} \\
	  \textsc{black} & & 
	\begin{tabular}{c}\includegraphics[trim = 0cm 0cm 0cm 0cm, width=0.22\columnwidth]{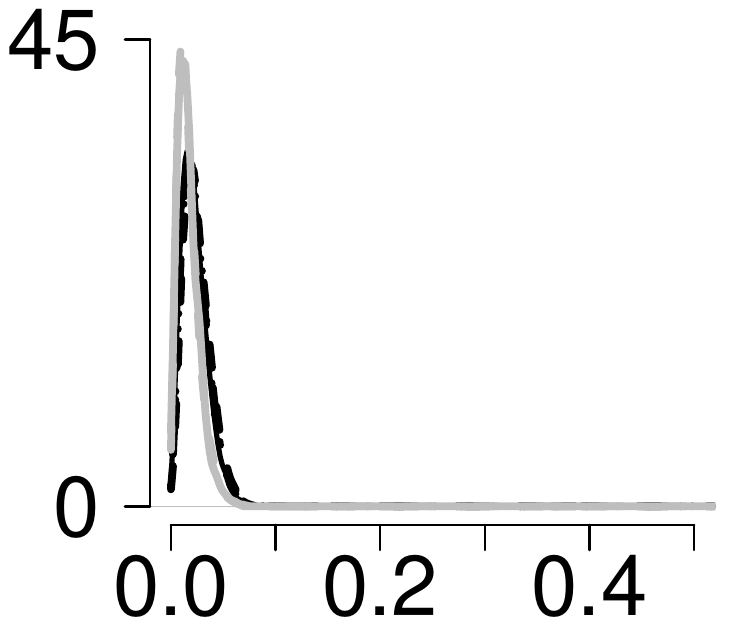}\end{tabular} & 
	\begin{tabular}{c}\includegraphics[trim = 0cm 0cm 0cm 0cm, width=0.22\columnwidth]{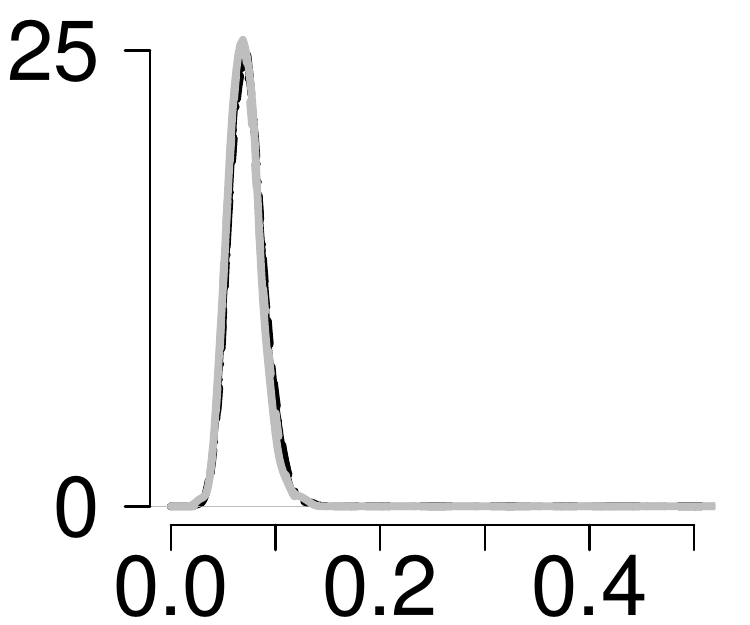}\end{tabular} & 
	\begin{tabular}{c}\includegraphics[trim = 0cm 0cm 0cm 0cm, width=0.22\columnwidth]{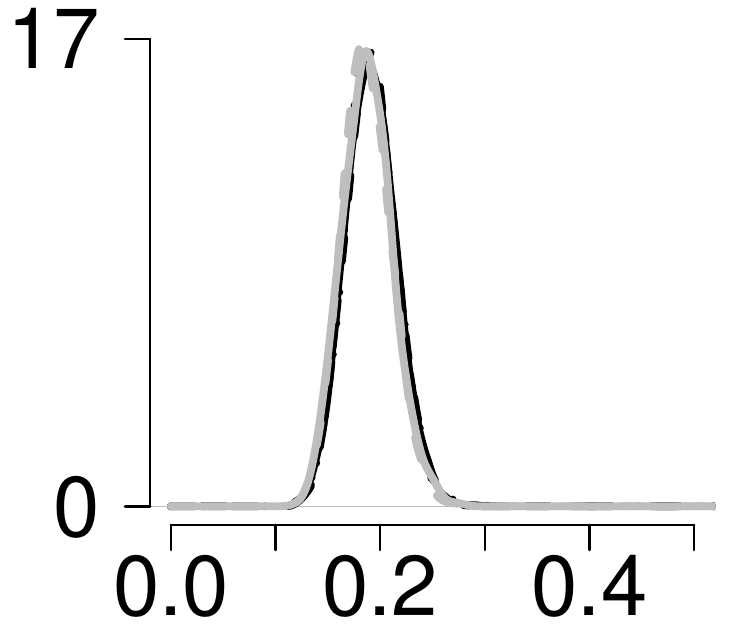}\end{tabular} &
	\begin{tabular}{c}\includegraphics[trim = 0cm 0cm 0cm 0cm, width=0.22\columnwidth]{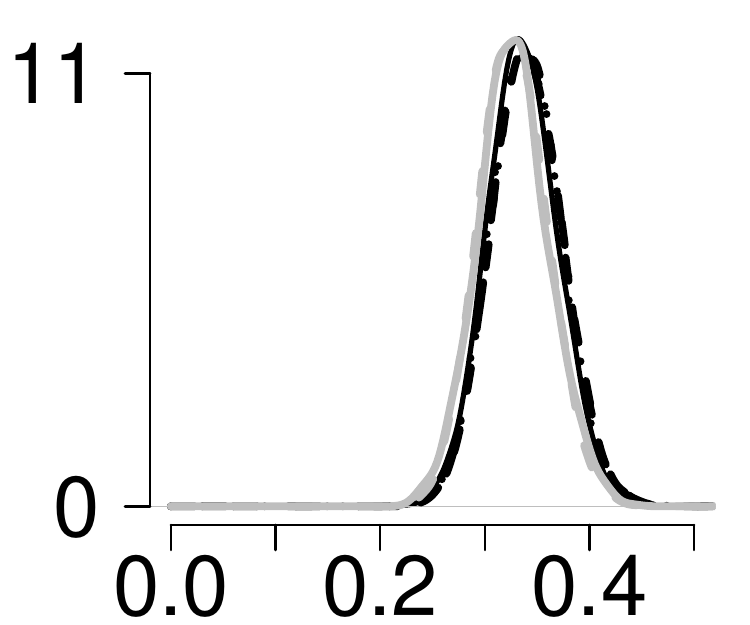}\end{tabular} \\ 	\textsc{white} & & 
	\begin{tabular}{c}\includegraphics[trim = 0cm 0cm 0cm 0cm, width=0.22\columnwidth]{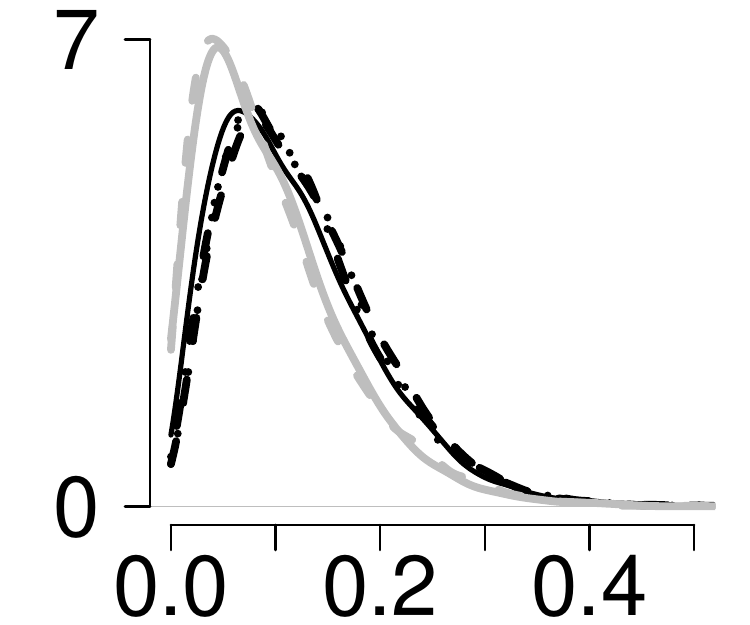}\end{tabular} & 
	\begin{tabular}{c}\includegraphics[trim = 0cm 0cm 0cm 0cm, width=0.22\columnwidth]{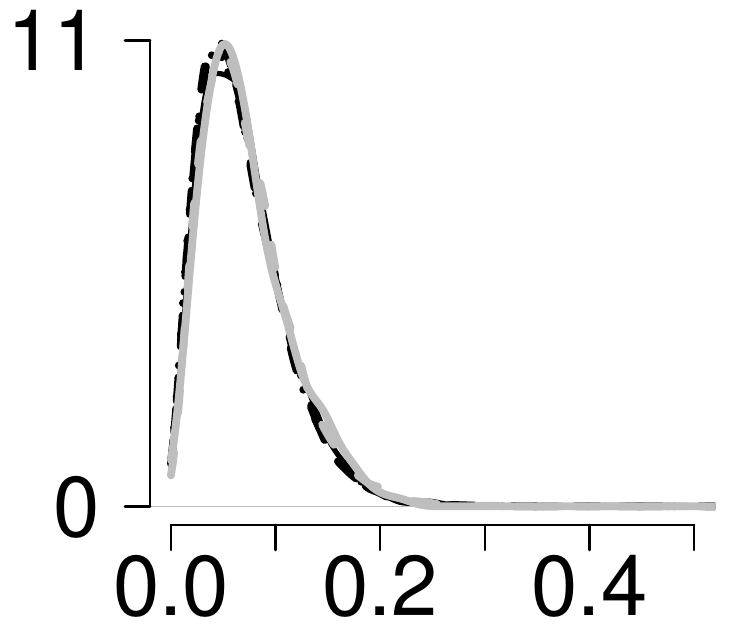}\end{tabular} & 
	\begin{tabular}{c}\includegraphics[trim = 0cm 0cm 0cm 0cm, width=0.22\columnwidth]{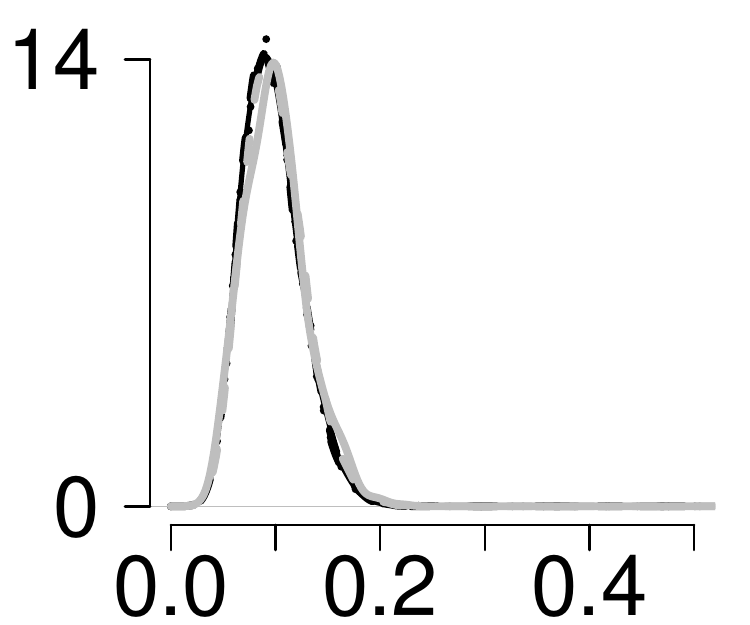}\end{tabular} &
	\begin{tabular}{c}\includegraphics[trim = 0cm 0cm 0cm 0cm, width=0.22\columnwidth]{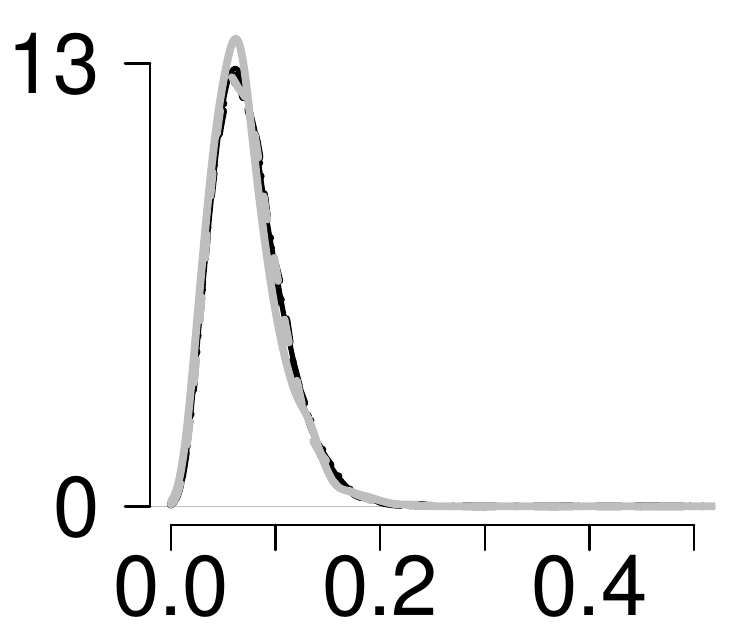}\end{tabular} \\ 	\textsc{other} & & 
  \begin{tabular}{c}\includegraphics[trim = 0cm 1cm 0cm 0cm, width=0.22\columnwidth]{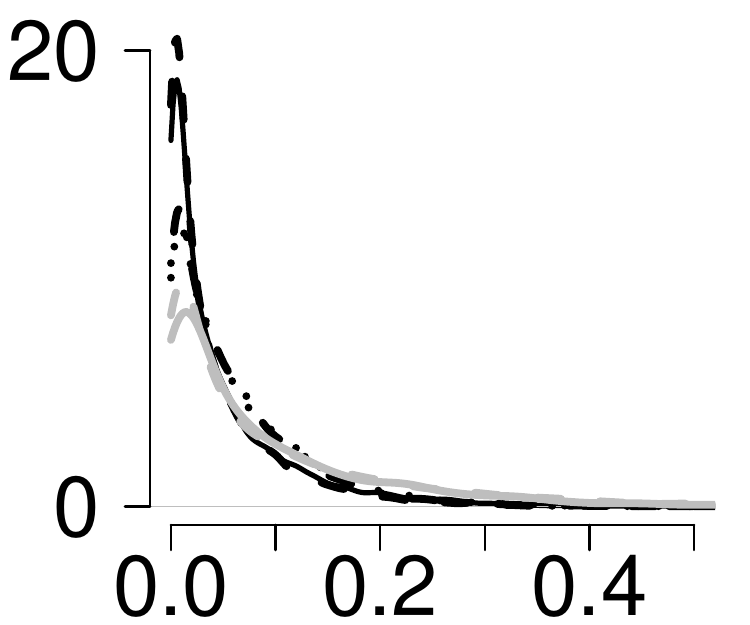}\end{tabular} & 
	\begin{tabular}{c}\includegraphics[trim = 0cm 1cm 0cm 0cm, width=0.22\columnwidth]{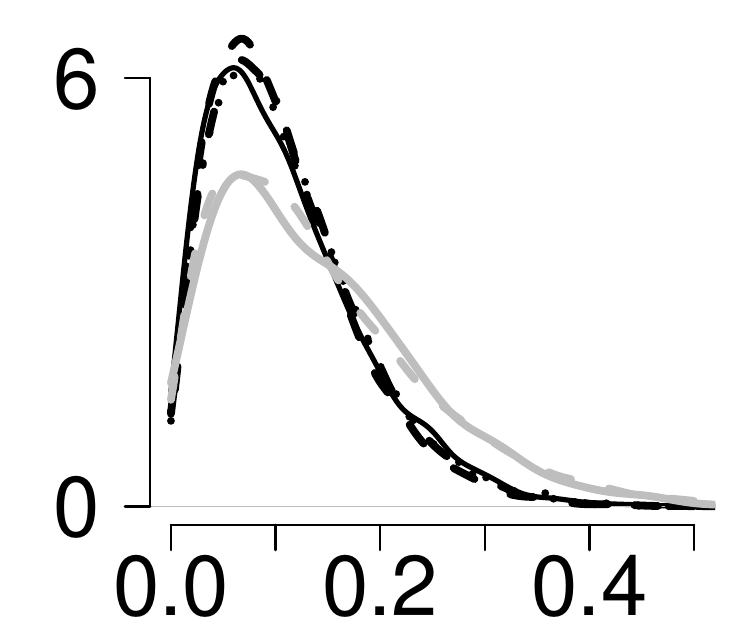}\end{tabular} & 
	\begin{tabular}{c}\includegraphics[trim = 0cm 1cm 0cm 0cm, width=0.22\columnwidth]{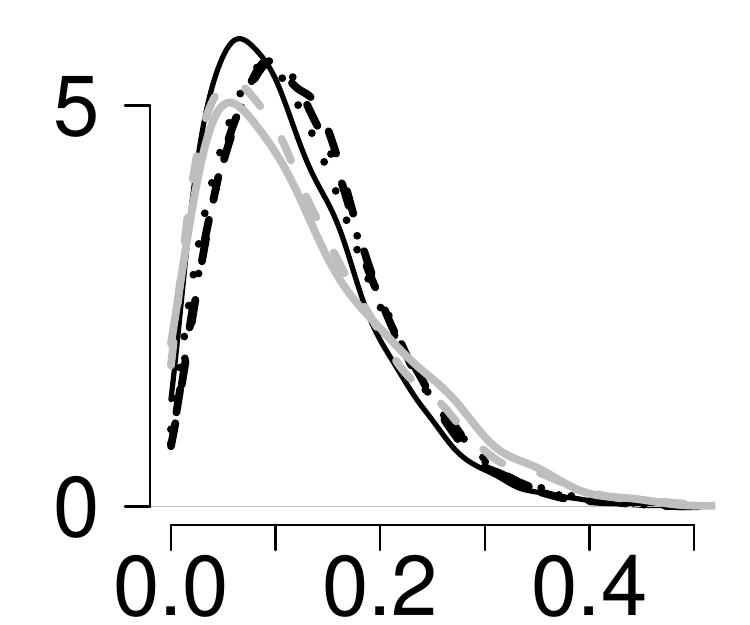}\end{tabular} &
	\begin{tabular}{c}\includegraphics[trim = 0cm 1cm 0cm 0cm, width=0.22\columnwidth]{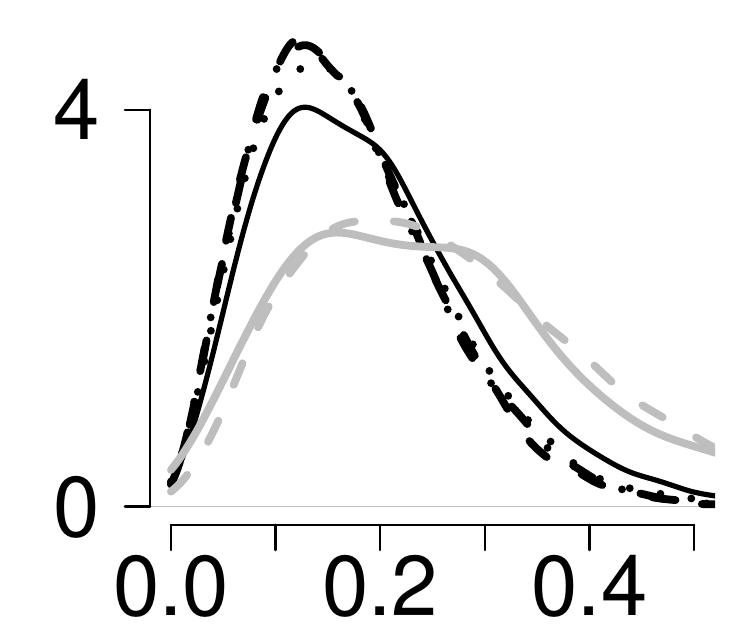}\end{tabular} \\
\end{tabular*}
\caption{Posterior distributions of proportion of females with diabetes among combinations of age and race under different missingness mechanisms.  Model 0: full sequential additive nonignorability, black solid line; model 1: main effects, black dashed line; model 2: order-invariant, black dot-dashed line; model 3: only-directly dependent, black dotted line; model 4: not-directly dependent, gray dashed line; model 5: ignorable model, gray solid line.} \label{t:female_diabetes}
\end{minipage}
\end{figure}

\begin{figure}[t]
  \centering
  \begin{minipage}[b]{1\textwidth}
 \def\~{\hphantom{0}}
  \begin{tabular*}{\columnwidth}{c@{\extracolsep{\fill}}c@{\extracolsep{\fill}}c@{\extracolsep{\fill}}c@{\extracolsep{\fill}}c@{\extracolsep{\fill}}c@{\extracolsep{\fill}}c@{\extracolsep{\fill}}}
 \multicolumn{3}{c}{\textsc{male}}  & & \multicolumn{3}{c}{\textsc{female}} \\
  \textsc{black} & \textsc{white} & \textsc{other} & & \textsc{black} & \textsc{white} & \textsc{other} \\
		\begin{tabular}{c}\includegraphics[trim = 0cm 1cm 0cm 1cm, width=0.16\columnwidth]{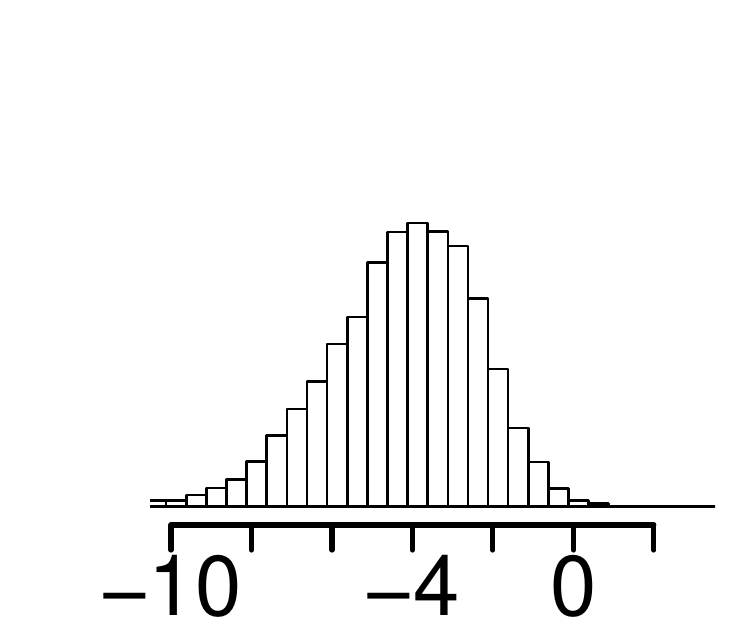}\end{tabular} & 
	\begin{tabular}{c}\includegraphics[trim = 0cm 1cm 0cm 1cm, width=0.16\columnwidth]{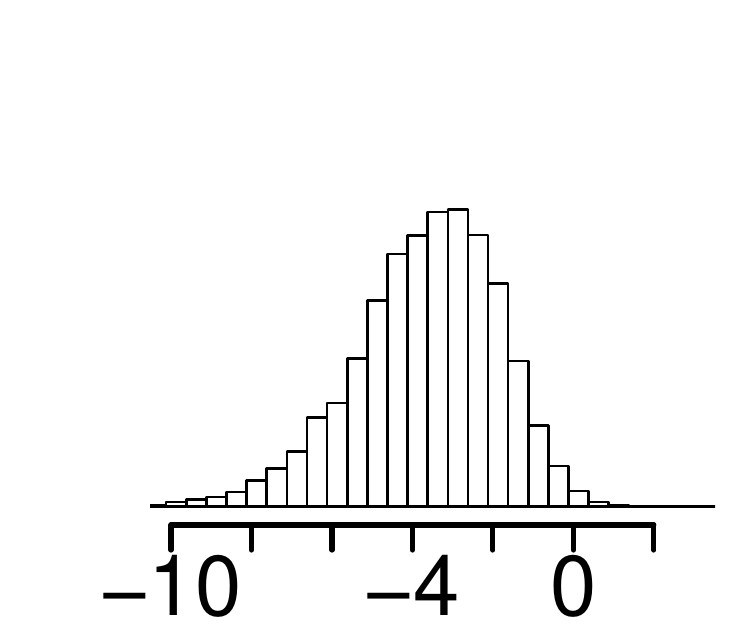}\end{tabular} & 
	\begin{tabular}{c}\includegraphics[trim = 0cm 1cm 0cm 1cm, width=0.16\columnwidth]{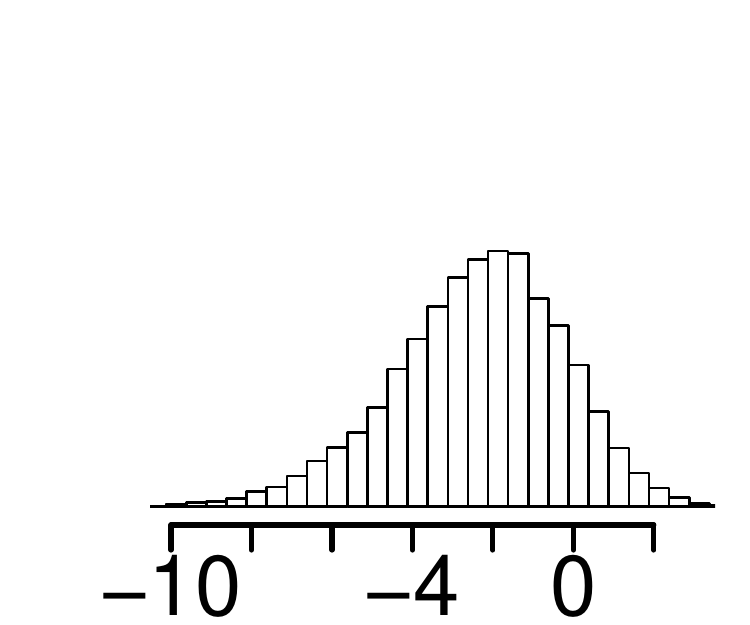}\end{tabular} & & 
	\begin{tabular}{c}\includegraphics[trim = 0cm 1cm 0cm 1cm, width=0.16\columnwidth]{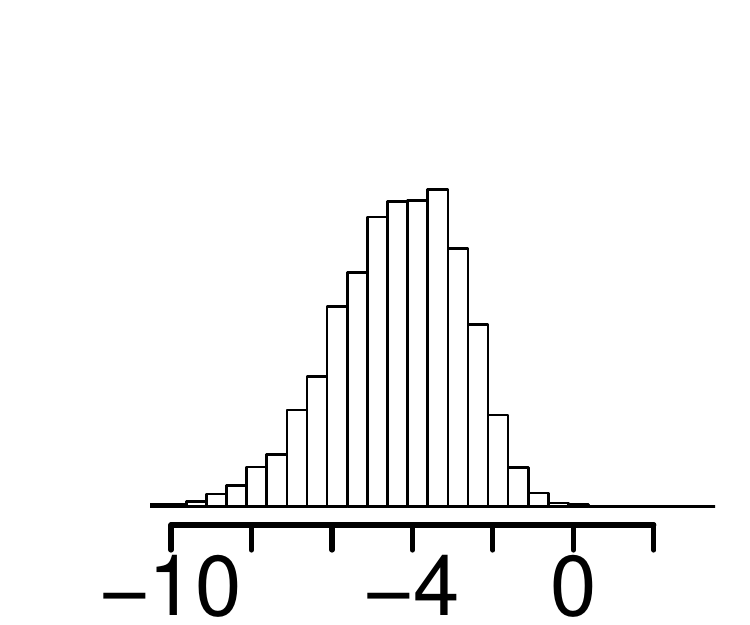}\end{tabular} & 
	\begin{tabular}{c}\includegraphics[trim = 0cm 1cm 0cm 1cm, width=0.16\columnwidth]{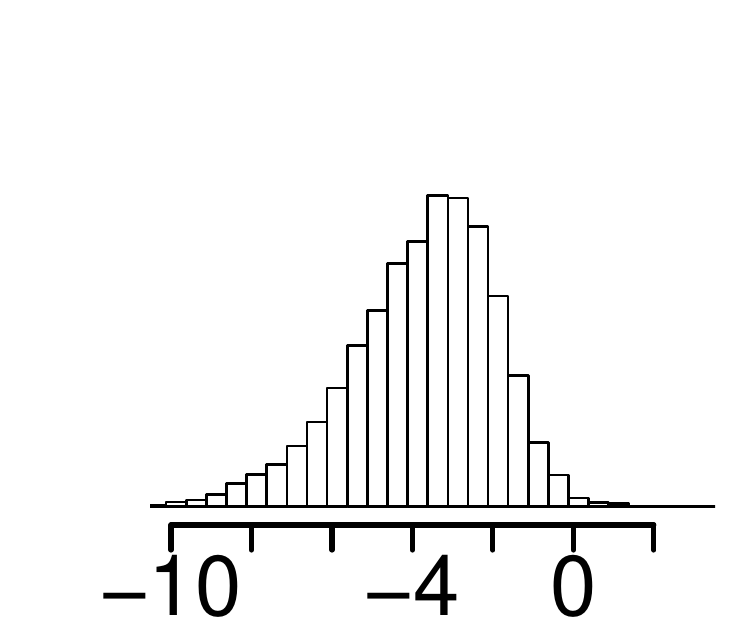}\end{tabular} & 
	\begin{tabular}{c}\includegraphics[trim = 0cm 1cm 0cm 1cm, width=0.16\columnwidth]{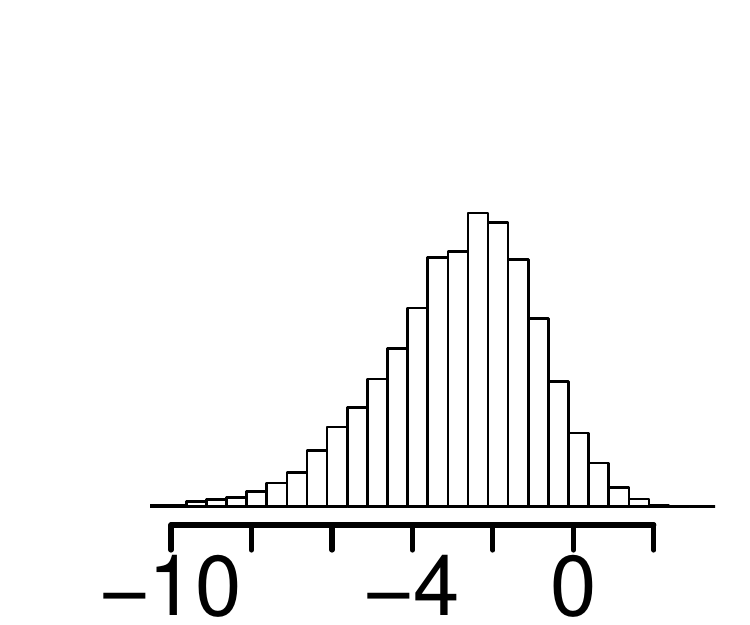}\end{tabular}\\ 
\end{tabular*}
\caption{Posterior distributions of log-odds ratios of nonresponse in age question for age \textsc{65+} versus baseline \textsc{20--34}, for combinations of race and sex.} \label{t:oddsM1}
\end{minipage}
\end{figure}

\begin{figure}[h]
  \centering
  \begin{minipage}[b]{1\textwidth}
 \def\~{\hphantom{0}}
  \begin{tabular*}{\columnwidth}{c@{\extracolsep{\fill}}c@{\extracolsep{\fill}}c@{\extracolsep{\fill}}c@{\extracolsep{\fill}}c@{\extracolsep{\fill}}}
\multicolumn{2}{c}{\textsc{male}}  & & \multicolumn{2}{c}{\textsc{female}} \\
 \textsc{white} & \textsc{other} & & \textsc{white} & \textsc{other} \\
	\begin{tabular}{c}\includegraphics[trim = 0cm 1cm 0cm 1cm, width=0.24\columnwidth]{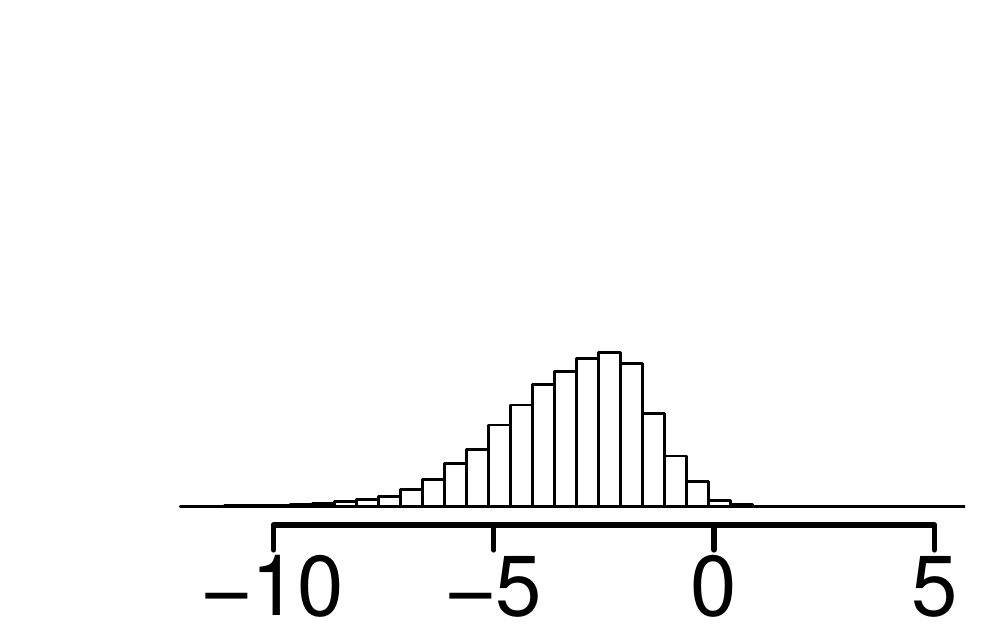}\end{tabular} & 
	\begin{tabular}{c}\includegraphics[trim = 0cm 1cm 0cm 1cm, width=0.24\columnwidth]{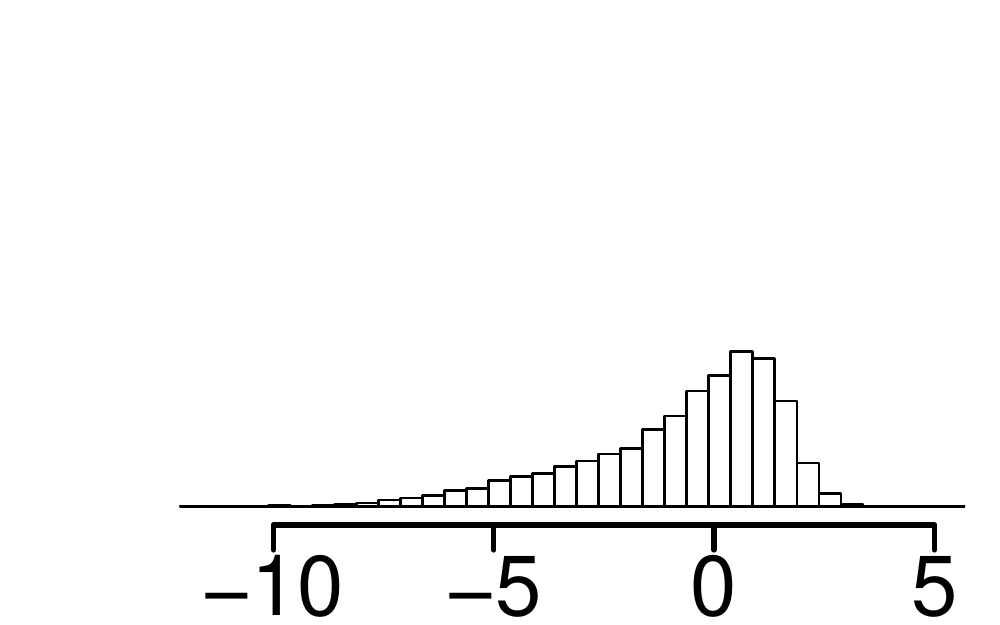}\end{tabular} & & 
	\begin{tabular}{c}\includegraphics[trim = 0cm 1cm 0cm 1cm, width=0.24\columnwidth]{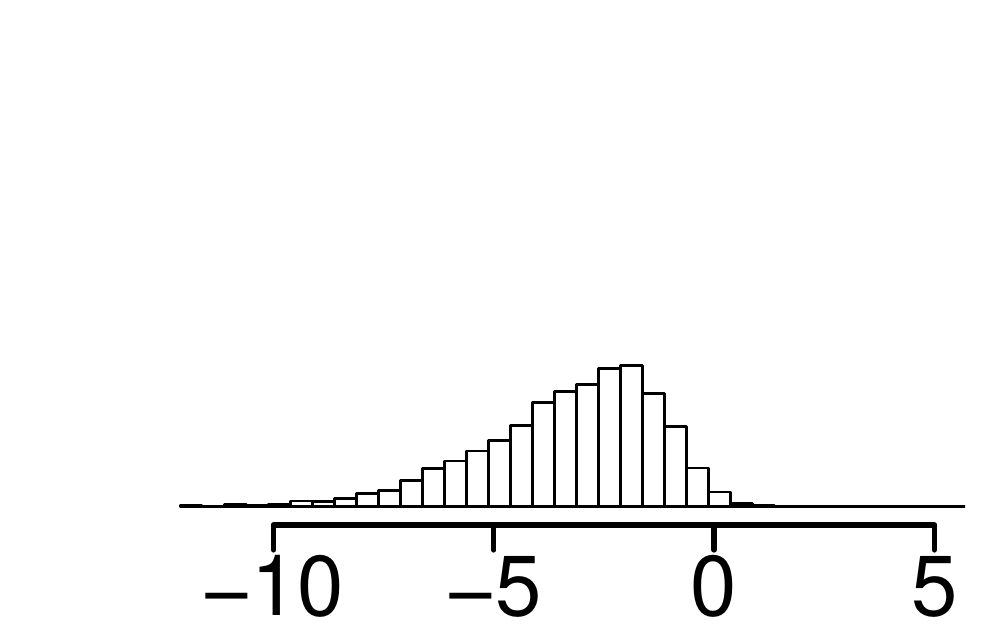}\end{tabular} & 
	\begin{tabular}{c}\includegraphics[trim = 0cm 1cm 0cm 1cm, width=0.24\columnwidth]{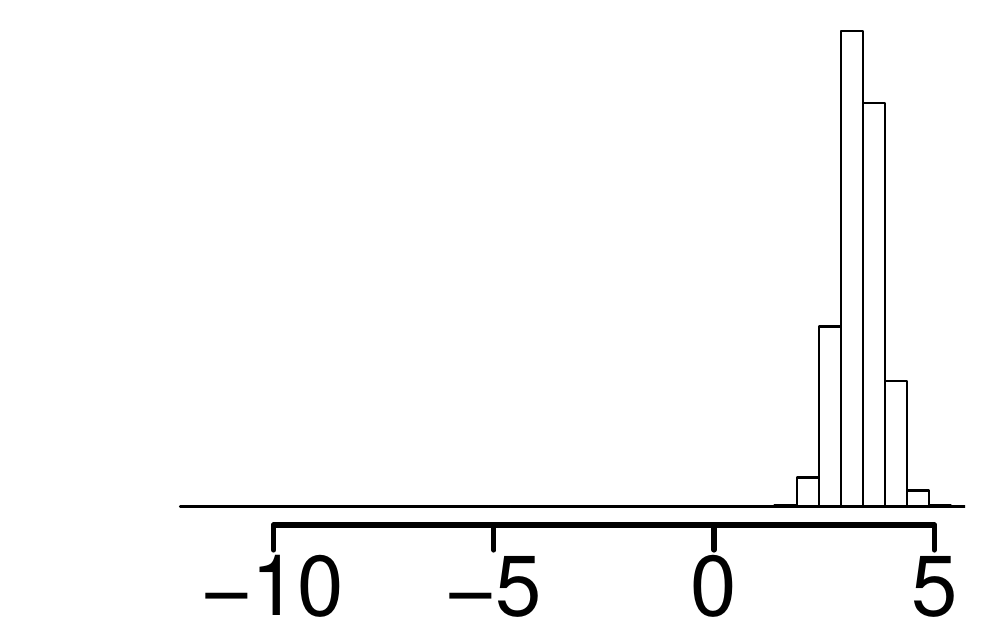}\end{tabular}\\ 
\end{tabular*}
\caption{Posterior distributions of log-odds ratios of nonresponse in race question for races \textsc{white} and \textsc{other} versus baseline \textsc{black}, among males and females.} \label{t:oddsM2}
\end{minipage}
\end{figure}

In Fig. \ref{t:female_diabetes} we present the posterior distributions of $\theta(y_1,y_2,\textsc{female})$, that is, the diabetes prevalence among females of age $y_1$ and race $y_2$, under the six different missingness mechanisms of Table \ref{t:models}. The posteriors under models 0--3 are very similar, all of which encode a direct dependence of the nonresponse on the corresponding study variables.  On the other hand, models 4 and 5, which exclude a direct dependence of the missingness mechanism on the study variables, also lead to nearly the same posterior distributions.  
This indicates that the differences obtained between assuming ignorability and sequential additive nonignorability are mainly due to the direct dependence of the nonresponse on the study variables.  Namely, for these data the most relevant feature of the sequentially additive nonignorable mechamism is represented by  submodel 3.  Figures \ref{t:oddsM1}  and  \ref{t:oddsM2} make this evident, as we explain below. The posterior distributions of $\theta(y_1,y_2,\textsc{male})$ were not very sensitive to the missingness mechanism so we omit them.

\setlength{\tabcolsep}{0em}

In Fig. \ref{t:oddsM1} we present the posterior distributions of the log-odds ratios $\beta_1(\textsc{65+},y_2,y_3)$ in the full model for race $y_2$ and sex $y_3$.  The analogous posteriors for the age categories \textsc{35--49} and \textsc{50--64} are very similar, so we omit them.  Most of the mass of each of these posteriors is below zero, indicating that in this illustration the odds of nonresponse in the variable age when the respondent is \textsc{65+} are likely to be much smaller than when he or she is \textsc{20--34}, which is the baseline. This indicates a strong negative association between the variable age and its nonresponse.

In Fig. \ref{t:oddsM2} we present the posterior distributions of the log-odds ratios $\beta_2(y_2,y_3)$ in the full model for race $y_2$ and sex $y_3$.  The posteriors for both $\beta_2(\textsc{white},\textsc{male})$ and $\beta_2(\textsc{white},\textsc{female})$ have masses mostly below zero, indicating that in this illustration the odds of nonresponse in the variable race for white respondents are likely to be much smaller than for black respondents.  The posterior of $\beta_2(\textsc{other},\textsc{male})$ is centered around zero, indicating that the odds of nonresponse in the race variable among males is similar for blacks and for people of other race.  On the other hand, the posterior of $\beta_2(\textsc{other},\textsc{female})$ is fully concentrated above zero, indicating that the odds of nonresponse in the race variable are higher for females of other race compared with black females.

As mentioned throughout the article, the assumptions encoded by the sequentially additive nonignorable missingness mechanism rely on an ordering of the variables, and changing this order leads to different missingness mechanisms, with their full versions enjoying nonparametric identification.  This property leads to a natural way of performing global or local sensitivity analyses \citep[e.g.,][]{Scharfstein18} by obtaining inferences under completely or partially different orderings of the variables, respectively.  

The main analysis that we presented relied on the order in which the variables were collected, but we also performed our analyses changing the order of the age and race variables.  The order of the sex variable is not relevant as it is fully observed, nor is the order of the diabetes variable since its missingness is assumed to be ignorable.  By construction, models 2, 3 and 5 do not depend on the order of age and race.  Changing the order of age and race in models 0, 1 and 4 could potentially lead to very different results, but with these data we found that the posterior distributions of the proportions of people with diabetes were virtually the same as in Fig. \ref{t:female_diabetes}.  This finding is intuitive given that the results in Fig. \ref{t:female_diabetes} are  essentially the same for models 0--3, and models 2 and 3 do not depend on the ordering of age and race.  We also replicated all of our results using the probit link, using the data augmentation scheme of \cite{AlbertChib93}, and found that the results were virtually the same as using the logit link.  

The results of this illustration indicate the existence of direct dependence of the nonresponse on the values of the items or variables being measured.  This direct dependence can be quantified thanks to the availability of auxiliary marginal information, in this case coming from a census, and thanks to the theoretical results presented in this article, which permits us to identify missingness mechanisms where this direct dependence occurs.

\section{Final remarks and future work}

The implementation that we proposed in Section \ref{ss:likelihood} relied on a likelihood function where the density of the distribution of $Y$, $f(y\mid \kappa)$, is parameterized explicitly in terms of features for which we have auxiliary marginal information.  This is not restrictive in the case of categorical variables, as in our application, but it might become so for continuous variables.  For example, if one is willing to assume that $Y$ is multivariate Gaussian, $\kappa$ would represent means, variances and covariances. We could then use auxiliary marginal information on some of those parameters, but information beyond the first two moments could not be used in this case.  Our identification results, however, are general enough to allow an arbitrary modeling of the distribution of $Y$.  Thus, a natural future avenue of research is to combine nonparametric or highly flexible models of the distribution of $Y$ with sequentially additive nonignorable missingness mechanisms.   For example, we could incorporate information on summaries of distributions into nonparametric Bayesian models using the approach of \cite{Kessler15}.

\section*{Acknowledgement}
The authors were supported by two grants from the U.S.A. National Science Foundation. 

\appendix

\section*{Appendix 1}\label{a:Iproj}
\subsection*{$\fdiv$-projections}

The monograph of \cite{LieseVajda87} contains a compendium of results on $\fdiv$-projections, some of which we present here for completeness.  We let $P$ and $Q$ be two probability distributions,  $\mathcal{P}$ be a set of probability distributions, and $I_\fdiv(P,Q)=\int \fdiv(dP/dQ)dQ$ be the $\fdiv$-divergence of $P$ and $Q$.  We start with Propositions 8.2 and 8.5 of \cite{LieseVajda87} which address the uniqueness and existence of $\fdiv$-projections.

\begin{theorem}[Uniqueness of $\fdiv$-projections]\label{LV:uniqueness}
Let $\mathcal{P}$ be convex, where each $P\in \mathcal{P}$ is dominated by $Q$.  Let $\fdiv$ be strictly convex at every point of $(0,\infty)$ and let $\inf_{P\in \mathcal{P}}I_\fdiv(P,Q)<\infty$.  Then there exists at most one $\fdiv$-projection of $Q$ onto $\mathcal{P}$.
\end{theorem}

\begin{theorem}[Existence of $\fdiv$-projections]\label{LV:existence}
Let $\mathcal{P}$ be convex and closed in variational distance.  Let $\lim_{z\rightarrow \infty}\fdiv(z)/z=\infty$.  Then there exists an $\fdiv$-projection of $Q$ onto $\mathcal{P}$.
\end{theorem}

The following result is simply a version of Theorem 8.20 in \cite{LieseVajda87} after combining it with their Lemma 8.19.

\begin{theorem}[Characterization of $\fdiv$-projections]\label{LV:characterization}
Let $\fdiv$ be a strictly convex differentiable function that satisfies Assumption \ref{f-growth}.  Let $\fdiv^*(z)=z\fdiv(1/z)$ for $z\in(0,\infty)$ also satisfy Assumption \ref{f-growth}.  Let $\mathcal{P}(\mathcal{U})$ be the set of probability distributions where each $P\in \mathcal{P}(\mathcal{U})$ has the same known finite value of $\int u dP$ for each function $u$ in a set $\mathcal{U}$.  Then $\langle\mathcal{U}\rangle\subset L_1(P)$ for each $P\in \mathcal{P}(\mathcal{U})$ and 
\begin{enumerate}
\item $P^*\in \mathcal{P}(\mathcal{U})$ is the $\fdiv$-projection of $Q$ onto $\mathcal{P}(\mathcal{U})$ if $\fdiv'(dP^*/dQ)\in \langle\mathcal{U}\rangle$.
\item If $P^*$ is the $\fdiv$-projection of $Q$ onto $\mathcal{P}(\mathcal{U})$ then $\fdiv'(dP^*/dQ)\in \overbar{\langle\mathcal{U}\rangle}$, where the closure is in $L_1(P^*)$.
\end{enumerate}
\end{theorem}

Similar results to this characterization can be found in \cite{Csiszar75}, \cite{Ruschendorf84}, and \cite{BroniatowskiKeziou06}.

\section*{Appendix 2}\label{a:proofs}
\subsection*{Proofs}

\begin{proof}[of Theorem \ref{th:1}]
The $\fdiv_\lambda$ function is differentiable by construction, and it is strictly convex given that its derivative $\fdiv_\lambda'(z) = \lambda[z/(c+z)]$ is monotonically increasing in $(0,\infty)$.  We can see that $\lim_{z\rightarrow \infty}\fdiv_\lambda^*(z)/z=0$, which is a sufficient condition for $\fdiv_\lambda^*$ to satisfy Assumption \ref{f-growth} \citep[][p. 171]{LieseVajda87}.
Now, we can rewrite $f(x,y\mid M=1)=f(x,y\mid M=0)\varphi\{f(M=1\mid x,y)\}$, where $\varphi(z)=cz/(1-z)$, with $c=(1-\pi)/\pi$.  We also assumed $f(M=1\mid x, y) = \lambda^{-1}\{\alpha(x)+\beta(y)\}$.  This means that $dF_1/dF_0(x,y)=\varphi[\lambda^{-1}\{\alpha(x)+\beta(y)\}]$, where $F_m$ denotes the distribution of $X,Y\mid M=m$, for $m=0,1$.  We then obtain $\fdiv_\lambda'\{dF_1/dF_0(x,y)\}=\alpha(x)+\beta(y)\in\langle L_1\{f(x\mid M=1)\}\cup\mathcal{U}\rangle$.  Theorem \ref{LV:characterization} then implies that $F_1$ is the $\fdiv_\lambda$-projection of $F_0$ onto the set of distributions with $X$-marginal given by $f(x\mid M=1)$ and with expected values of each $u\in\mathcal{U}$ matching those determined by the auxiliary marginal information, $\{\int u(y) f(y\mid M=1)\mu(dy)\}_{u\in\mathcal{U}}$.  Theorem \ref{th:1} simply states this result in terms of densities of $F_1$ and $F_0$.
\end{proof}

\begin{proof}[of Theorem \ref{th:NPS_AN}]
Let us denote by $G_m$ and $H_m$ the distributions with densities $g(x,y\mid M=m)$ and $h(x,y\mid M=m)$, respectively, $m=0,1$.  Let us denote by $P^X$ and  $P^Y$ the $X$- and $Y$-marginals of a distribution $P$.  By assumption, $G_1$ is the $\fdiv_\lambda$-projection of $H_0$ onto the set of distributions $\mathcal{P}$ where each element $P$ is such that $\int v dP^X=\int v dH_1^X$ and $\int u dP^Y=\int u dH_1^Y$, for all $v\in L_1(H_1^X)$ and all $u\in\mathcal{U}$.  To guarantee the existence of $G_1$, we note that $\mathcal{P}$ is convex as convex combinations of distributions that satisfy the constraints also satisfy the constraints, and is closed in variational distance since it is the solution set of a number of equations given by the constraints.  Furthermore, by L'H\^opital's rule we find that $\lim_{z\rightarrow \infty}\fdiv_\lambda(z)/z=\lim_{z\rightarrow \infty}\lambda\{z/(c+z)\}=\infty$.  Therefore Theorem \ref{LV:existence} guarantees the existence of $G_1$, and its uniqueness is obtained from Theorem \ref{LV:uniqueness} since $\fdiv_\lambda$ is strictly convex given that its derivative $\fdiv_\lambda'(z) = \lambda[z/(c+z)]$ is monotonically increasing in $(0,\infty)$.

Similarly as in the proof of Theorem \ref{th:1}, we find that $\fdiv_\lambda$ and $\fdiv_\lambda^*$ satisfy the conditions of Theorem \ref{LV:characterization}.  Therefore Theorem \ref{LV:characterization} implies that $\fdiv_\lambda'(dG_1/dH_0)\in \overbar{\langle L_1(H_1^X)\cup \mathcal{U}\rangle}$, where the functions in $L_1(H_1^X)$ are constant in $y$ and the functions in $\mathcal{U}$ are constant in $x$.  Note that if $\fdiv_\lambda'(dG_1/dH_0)\in\langle L_1(H_1^X)\cup \mathcal{U}\rangle$, then it can be written as $\fdiv_\lambda'(dG_1/dH_0)=\alpha(x)+\beta(y)$ where $\alpha \in L_1\{H_1^X\}$ and $\beta\in\langle\mathcal{U}\rangle$, but if it is a limit point outside of that set then it can be arbitrarily approximated  by functions of such form.  Now, from the definition of $g(x,y,m)$, we find that $g(M=1\mid x,y)=dG_1/dH_0(x,y)/\{c+dG_1/dH_0(x,y)\}$, since $g(x,y\mid M=1)/h(x,y\mid M=0)=dG_1/dH_0(x,y)$.  Therefore, we conclude $\lambda[g(M=1\mid x,y)]=\fdiv_\lambda'[dG_1/dH_0(x,y)]\in \overbar{\langle L_1(H_1^X)\cup \mathcal{U}\rangle}$.  Now, $g(x,y^*)=h(x,y^*)$ because by construction $g(x,y,M=0)=h(x,y,M=0)$, and $g(x,M=1)=\pi g(x\mid M=1)=\pi h(x\mid M=1)$ from how we construct $g(x,y\mid M=1)$ as an $\fdiv_\lambda$-projection.  Finally, we also have that $\int u(y) g(y)\mu(dy)=\int u(y) h(y)\mu(dy)$ because $g(M=1)=h(M=1)$, and $\int u(y) g(y\mid M=m)\mu(dy)=\int u(y) h(y\mid M=m)\mu(dy)$ for all $u\in\mathcal{U}$, when $m=1$ based on how we construct $g(x,y\mid M=1)$ as an $\fdiv_\lambda$-projection, and when $m=0$ given that $g(y\mid M=0)=h(y\mid M=0)$ by construction of $g(x,y,m)$.
\end{proof}

\begin{proof}[of Theorem \ref{th:SAN_ident}]
1. Analogously to the proof of Theorem \ref{th:1}, we first find that all $\fdiv_{\lambda,j}$ and $\fdiv_{\lambda,j}^*$ satisfy the conditions required by Theorem \ref{LV:characterization}, and we obtain $\fdiv_{\lambda,j}'\{dF_{1,j}/dF_{0,j}(x,y)\}=\alpha_j(x,y^*_{<j},y_{>j})+\beta_j(y_{\geq j})\in\langle L_1\{f(x,y^*_{<j},y_{>j}\mid M=1)\}\cup\mathcal{U}_{\geq j}\rangle$, where $F_{m,j}$ is the distribution with density $f(x,y^*_{<j},y_{\geq j}\mid M_j=m)$, $m=0,1$.  Theorem \ref{LV:characterization} then implies that $F_{1,j}$ is the $\fdiv_{\lambda,j}$-projection of $F_{0,j}$ onto the set of distributions with marginal determined by $f(x,y^*_{<j},y_{>j}\mid M=1)$ and with expected values of each $u\in\mathcal{U}_{\geq j}$ matching those determined by the auxiliary marginal information, $\{E[u(Y_{\geq j})\mid M_j=1]; u\in \mathcal{U}_{\geq j}\}$.  

2. Part 1 of this theorem guarantees that the true density $f(x,y^*_{<j},y_{\geq j}\mid M_j=1)$ can be recovered 
from $f(x,y^*_{<j},y_{\geq j}\mid M_j=0)$, $f(x,y^*_{<j},y_{>j}\mid M_j=1)$ and $\{E[u(Y_{\geq j})\mid M_j=1]; u\in \mathcal{U}_{\geq j}\}$, for each $j=1,\ldots,p$.  Algorithm \ref{al1} implements the sequence of projections justified by Part 1.  For $j=1$ we obtain $f(x,y)$ from the algorithm's substep d, and the missingness mechanism is obtained as $f(m\mid x,y)=\prod_{j=1}^p f(m_j\mid x,y^*_{<j},y_{\geq j})$, where $f(m_j\mid x,y^*_{<j},y_{\geq j})$ is obtained in step $j$ of the algorithm.
\end{proof}

\begin{proof}[of Theorem \ref{th:SAN_NPS}]
1. In the construction of $g(x,y,m)$, we find $g(x,y^*_{<j},y_{\geq j}\mid M_j=1)$ as the $\fdiv_{\lambda,j}$-projection of $g(x,y^*_{<j},y_{\geq j}\mid M_j=0)$ onto the set of distributions that match the marginal $g(x,y^*_{<j},y_{> j}\mid M_j=1)$ and the expectations $\{E_g[u(Y_{\geq j})\mid M_j=1]; u\in \mathcal{U}_{\geq j}\}$, with $\fdiv_{\lambda,j}(z) = \int_{0}^z\lambda[v/(c_j+v)]dv$, $c_j=(1-\pi_j)/\pi_j$.  
Confirming the conditions required by Theorems \ref{LV:existence} and \ref{LV:uniqueness} to guarantee the existence and uniqueness of each $g(x,y^*_{<j},y_{\geq j}\mid M_j=1)$ is analogous as in the proof of Theorem \ref{th:NPS_AN}, so we omit it.  
Finding that $\fdiv_{\lambda,j}$ and $\fdiv_{\lambda,j}^*$ satisfy the conditions of Theorem \ref{LV:characterization} is also analogous as in the proof of Theorem \ref{th:1}, so we also omit it.  Denoting by $G_{m,j}$ the distribution with density $g(x,y^*_{<j},y_{\geq j}\mid M_j=m)$, $m=0,1$, Theorem \ref{LV:characterization} implies that $\fdiv_{\lambda,j}'(dG_{1,j}/dG_{0,j})\in \overbar{\langle L_1\{g(x,y^*_{<j},y_{> j}\mid M_j=1)\}\cup \mathcal{U}_{\geq j}\rangle}$, where the functions in $L_1\{g(x,y^*_{<j},y_{> j}\mid M_j=1)\}$ are constant in $y_j$ and the functions in $\mathcal{U}_{\geq j}$ are constant in $(x,y^*_{<j})$.  
Now, by construction we find that $g(M_j=1\mid x,y^*_{<j},y_{\geq j})=dG_{1,j}/dG_{0,j}(x,y^*_{<j},y_{\geq j})/\{c_j+dG_{1,j}/dG_{0,j}(x,y^*_{<j},y_{\geq j})\}$, since $g(x,y^*_{<j},y_{\geq j}\mid M_j=1)/g(x,y^*_{<j},y_{\geq j}\mid M_j=0)=dG_{1,j}/dG_{0,j}(x,y)$.  Therefore, we conclude $\lambda[g(M_j=1\mid x,y^*_{<j},y_{\geq j})]=\fdiv_{\lambda,j}'[dG_{1,j}/dG_{0,j}(x,y^*_{<j},y_{\geq j})]\in \overbar{\langle L_1\{g(x,y^*_{<j},y_{> j}\mid M_j=1)\}\cup \mathcal{U}_{\geq j}\rangle}$.  

2. We first show that the observed-data distribution implied by $g(x,y,m)$ is $h(x,y^*)$.  The result of Algorithm \ref{al1} is $g(x,y,m)= g(x,y)\prod_{j=1}^p g(m_j\mid x,y^*_{<j},y_{\geq j})$, which we need to integrate over the missing values $y_{m}$ according to a generic missingness pattern $m=(m_1,\dots,m_p)$.  To do this we sequentially integrate $g(x,y,m)$ over $y_j$ if $m_j=1$, $j=1,\dots,p$.  At each step we obtain $g(x,y^*_{\leq j},y_{>j},m_{>j})$ from $g(x,y_{<j}^*,y_{\geq j},m_{\geq j})$.  When $j=1$ this corresponds to obtaining  $g(x,y^*_{1},y_{>1},m_{>1})$ from $g(x,y,m)$, and when $j=p$ this corresponds to 
obtaining $g(x,y^*)$ from $g(x,y_{<p}^*,y_{p},m_{p})$.

Let us say that after having integrated over $(y_l: m_l=1, l<j)$ we  obtained $g(x,y_{<j}^*,y_{\geq j},m_{\geq j})$.  If $m_j=0$ then we do not have to integrate over $y_j$, and $g(x,y_{<j}^*,y_{\geq j},m_{\geq j})=g(x,y_{\leq j}^*,y_{> j},m_{> j})$ from the definition of $Y^*_j$.  If $m_j=1$ then we need to integrate over $y_j$.  From the construction in Algorithm \ref{al1}, $g(x,y_{<j}^*,y_{\geq j},m_{\geq j})=g(x,y_{<j}^*,y_{\geq j})\prod_{k\geq j}g(m_{k}\mid x,y_{<k}^*,y_{\geq k})$.  By definition, all $g(m_k\mid x,y^*_{<k},y_{\geq k})$ for $k>j$ do not depend on $y_j$ when $m_j=1$, and so we just need to integrate $g(x,y_{<j}^*,y_{\geq j})g(M_{j}=1\mid x,y_{<j}^*,y_{\geq j})=g(x,y^*_{<j},y_{\geq j}\mid M_j=1)\pi_j$ over $y_j$.  Now, $g(x,y^*_{<j},y_{\geq j}\mid M_j=1)$ is obtained as the $\fdiv_{\lambda,j}$-projection of $g(x,y^*_{<j},y_{\geq j}\mid M_j=0)$ onto the set constrained by the marginal $g(x,y^*_{<j},y_{> j}\mid M_j=1)$ and the expectations $\{E_g[u(Y_{\geq j})\mid M_j=1]; u\in \mathcal{U}_{\geq j}\}$.  By construction then $\int g(x,y^*_{<j},y_{\geq j}\mid M_j=1)\mu_j(dy_j)\pi_j=g(x,y^*_{<j},y_{>j}\mid M_j=1)\pi_j$, which can be written as $g(x,y^*_{\leq j},y_{>j})$ from the definition of $Y_j^*$ when $M_j=1$.  We then obtain for $m_j=1$, $g(x,y^*_{\leq j},y_{>j},m_{>j})=\int g(x,y_{<j}^*,y_{\geq j},m_{\geq j})\mu_j(dy_j)=g(x,y^*_{\leq j},y_{>j})\prod_{k> j}g(m_{k}\mid x,y_{<k}^*,y_{\geq k})$.  When $j=p$ we then  obtain $g(x,y^*_{\leq p},y_{>p},m_{>p})=g(x,y^*)=h(x,y^*)$.

Finally, to show that $\int u(y) g(y)\mu(dy)=\int u(y) h(y)\mu(dy)$ for all $u\in\mathcal{U}$, note that from the output of Algorithm \ref{al1} $g(x,y)=g(x,y\mid M_1=1)\pi_1+g(x,y\mid M_1=0)(1-\pi_1)$ from substep d. of the algorithm's last step.  Here $g(x,y\mid M_1=1)$ is the $\fdiv_{\lambda,1}$-projection of $g(x,y\mid M_1=0)$ onto the set with marginal given by $g(x,y_{>1}\mid M_1=1)$ and expectations given by $\{E_g[u(Y)\mid M_1=1]; u\in \mathcal{U}\}$, as derived from steps a. and b. of the algorithm's last step.  We	therefore obtain that $\int u(y) g(y)\mu(dy)=\int u(y) g(y\mid M_1=1)\mu(dy)\pi_1+\int u(y) g(y\mid M_1=0)\mu(dy)(1-\pi_1)=\int u(y) h(y)\mu(dy)$, given that $\int u(y) g(y\mid M_1=1)\mu(dy)=E_g[u(Y)\mid M_1=1]=\{E[u(Y)]-\int u(y)g(y,M_1=0)\mu(dy)\}/\pi_1$.  
\end{proof}

\bibliographystyle{APA}
\bibliography{biblio_bka}

\end{document}